# Iterative Secret Key Rate Adapting with Error Minimization for Continuous-Variable Quantum Key Distribution

Laszlo Gyongyosi

[1] Quantum Technologies Laboratory
Department of Networked Systems and Services
Budapest University of Technology and Economics
2 Magyar tudosok krt., Budapest, H-1117 Hungary
[2] MTA-BME Information Systems Research Group
Hungarian Academy of Sciences
7 Nador st., Budapest, H-1051 Hungary

gyongyosi@hit.bme.hu

**Abstract**

We define an iterative error-minimizing secret key adapting method for multicarrier CVQKD. A multicarrier CVQKD protocol uses Gaussian subcarrier quantum continuous variables (CVs) for the transmission. The proposed method allows for the parties to reach a given target secret key rate with minimized error rate through the Gaussian sub-channels by a sub-channel adaption procedure. The adaption algorithm iteratively determines the optimal transmit conditions to achieve the target secret key rate and the minimal error rate over the sub-channels. The solution requires no complex calculations or computational tools, allowing for easy implementation for experimental CVQKD scenarios.

# 1 Introduction

Continuous-variable quantum key distribution (CVQKD) provides an easily implementable solution to realize unconditional secure communication over the current telecommunication networks [10–22]. CVQKD does not require single-photon sources and detectors and can be implemented in an experimental scenario by standard devices [1], [9–26], [30–37]. In a CVQKD setting, the information is carried by a continuous-variable quantum state that is defined in the phase space via the position and momentum quadratures. In an experimental CVQKD scenario, the CV quantum states have a Gaussian random distribution, and the quantum channel between the sender (Alice) and receiver (Bob) is also Gaussian, because the presence of an eavesdropper (Eve) adds a white Gaussian noise into the transmission [19-41].

The CVQKD protocols have several smart properties. However, the relevant performance attributes of experimental CVQKD (i.e., secret key rates, transmission distances, tolerable excess noise, etc.) still require significant improvements. For this purpose, the multicarrier CVQKD has been recently introduced through the adaptive quadrature division modulation (AMQD) [2]. The multicarrier CVQKD scheme injects several additional degrees of freedom into the transmission, which are not available for a standard (single-carrier) CVQKD setting. In particular, the extra benefits and resources allow the realization of improved secret key rates and a higher amount of tolerable losses with unconditional security. These results also made possible to utilize several significant phenomena for CVQKD that are unavailable in a standard CVQKD protocol (such as single layer transmission [4], enhanced security thresholds [5], multidimensional manifold extraction [6], characterization of the subcarrier domain [7], adaptive quadrature detection and subchannel estimation techniques [8], extensive utilization of distribution statistics and random matrix formalism [9], and advanced statistical approaches for performance improvements [10], [41-42]).

In this work, we define an *iterative error-minimizing secret key adaption* method for multicarrier CVQKD. The proposed secret key adaption algorithm iteratively determines the optimal transmit conditions at a given target secret key rate to realize minimal error transmission over the sub-channels. At a given transmission rate of private classical information (private rate), the method determines and selects that sub-channel from the set of available sub-channels for the transmission of the quadratures, which sub-channel provides a minimal error rate. The secret key adaption successively utilizes private rate curves for the sub-channels. The curves also define an adaption region for each sub-channel. Particularly, the adaption region provides a base for the iterative, private rate increment method utilized by our secret key rate adaption algorithm. The iterative sub-channel selection procedure depends on the actual target private rate and the noise levels of the sub-channels. The scheme provably yields a minimized error rate transmission for all sub-channels while achieving the selected target secret key rate. We demonstrate the results through the framework of AMQD and also extend the results to the multiple-access multicarrier CVQKD.

This paper is organized as follows. In Section 2, preliminary findings are summarized. Section 3 discusses the iterative secret key adaption scheme. Section 4 extends the results to a mul-tiuser setting. Finally, Section 5 concludes the results. A numerical evidence is included in the Supplemental Information.

## 2 Preliminaries

In Section 2, the notations and basic terms are summarized. For further information, see the detailed descriptions of [2–10].

### 2.1 Multicarrier CVQKD

The following description assumes a single user, and the use of $n$ Gaussian sub-channels $\mathcal{N}_i$ for the transmission of the subcarriers, from which only $l$ sub-channels will carry valuable information.

In the single-carrier modulation scheme, the $j$-th input single-carrier state $|\varphi_j\rangle = |x_j + \mathrm{i}p_j\rangle$ is a Gaussian state in the phase space $\mathcal{S}$, with i.i.d. Gaussian random position and momentum quadratures

$$x_j \in \mathbb{N}\left(0, \sigma_{\omega_0}^2\right),\ p_j \in \mathbb{N}\left(0, \sigma_{\omega_0}^2\right), \tag{1}$$

where $\sigma_{\omega_0}^2$ is the modulation variance of the quadratures. In the multicarrier scenario, the information is carried by Gaussian subcarrier CVs, $|\phi_i\rangle = |x_i + \mathrm{i}p_i\rangle$, via quadratures

$$x_i \in \mathbb{N}\left(0, \sigma_{\omega}^2\right),\ p_i \in \mathbb{N}\left(0, \sigma_{\omega}^2\right), \tag{2}$$

where $\sigma_{\omega}^2$ is the modulation variance of the subcarrier quadratures, which are transmitted through a noisy Gaussian sub-channel $\mathcal{N}_i$. Precisely, each $\mathcal{N}_i$ Gaussian sub-channel is dedicated for the transmission of one Gaussian subcarrier CV from the $n$ subcarrier CVs. (*Note*: index $i$ refers to a subcarrier CV, index $j$ to a single-carrier CV, respectively.)

The single-carrier CV state $|\varphi_j\rangle$ in $\mathcal{S}$ can be modeled as a zero-mean, circular symmetric complex Gaussian random variable $z_j \in \mathcal{CN}\left(0, \sigma_{\omega_{z_j}}^2\right)$, with a variance

$$\sigma_{\omega_{z_j}}^2 = \mathbb{E}\left[|z_j|^2\right] = 2\sigma_{\omega_0}^2, \tag{3}$$

and with i.i.d. real and imaginary zero-mean Gaussian random components

$$\mathrm{Re}(z_j) \in \mathbb{N}\left(0, \sigma_{\omega_0}^2\right),\ \mathrm{Im}(z_j) \in \mathbb{N}\left(0, \sigma_{\omega_0}^2\right). \tag{4}$$

In the multicarrier CVQKD scenario, let $n$ be the number of Alice's input single-carrier Gaussian states. The $n$ input coherent states are modeled by an $n$-dimensional, zero-mean, circular symmetric complex random Gaussian vector

$$\mathbf{z} = \mathbf{x} + \mathrm{i}\mathbf{p} = \left(z_0, \ldots, z_{n-1}\right)^T \in \mathcal{CN}\left(0, \mathbf{K_z}\right), \tag{5}$$

where each $z_j$ is a zero-mean, circular symmetric complex Gaussian random variable

$$z_j \in \mathcal{CN}\left(0, \sigma^2_{\omega_{z_j}}\right), \; z_j = x_j + \mathrm{i}p_j. \tag{6}$$

In the first step of AMQD, Alice applies the inverse FFT (fast Fourier transform) operation to vector $\mathbf{z}$ (see (5)), which results in an $n$-dimensional zero-mean, circular symmetric complex Gaussian random vector $\mathbf{d}$, $\mathbf{d} \in \mathcal{CN}(0, \mathbf{K_d})$, $\mathbf{d} = (d_0, \dots, d_{n-1})^T$, precisely as

$$\mathbf{d} = F^{-1}(\mathbf{z}) = e^{\frac{\mathbf{d}^T \mathbf{A} \mathbf{A}^T \mathbf{d}}{2}} = e^{\frac{\sigma^2_{\omega_0}\left(d_0^2 + \dots + d_{n-1}^2\right)}{2}}, \tag{7}$$

where

$$d_i = x_{d_i} + \mathrm{i}p_{d_i}, \; d_i \in \mathcal{CN}\left(0, \sigma^2_{d_i}\right), \tag{8}$$

where $\sigma^2_{\omega_{d_i}} = \mathbb{E}\left[|d_i|^2\right] = 2\sigma^2_{\omega}$, thus the position and momentum quadratures of $|\phi_i\rangle$ are i.i.d. Gaussian random variables with a constant variance $\sigma^2_{\omega}$ for all $\mathcal{N}_i, i = 0, \dots, l-1$ sub-channels:

$$\mathrm{Re}(d_i) = x_{d_i} \in \mathbb{N}\left(0, \sigma^2_{\omega}\right), \; \mathrm{Im}(d_i) = p_{d_i} \in \mathbb{N}\left(0, \sigma^2_{\omega}\right), \tag{9}$$

where $\mathbf{K_d} = \mathbb{E}\left[\mathbf{d}\mathbf{d}^\dagger\right]$, $\mathbb{E}[\mathbf{d}] = \mathbb{E}\left[e^{\mathrm{i}\gamma}\mathbf{d}\right] = \mathbb{E}e^{\mathrm{i}\gamma}[\mathbf{d}]$, $\mathbb{E}\left[\mathbf{d}\mathbf{d}^T\right] = \mathbb{E}\left[e^{\mathrm{i}\gamma}\mathbf{d}\left(e^{\mathrm{i}\gamma}\mathbf{d}\right)^T\right] = \mathbb{E}e^{\mathrm{i}2\gamma}\left[\mathbf{d}\mathbf{d}^T\right]$ for any $\gamma \in [0, 2\pi]$.

The $\mathbf{T}(\mathcal{N})$ transmittance vector of $\mathcal{N}$ in the multicarrier transmission is

$$\mathbf{T}(\mathcal{N}) = \left[T_0(\mathcal{N}_0), \dots, T_{n-1}(\mathcal{N}_{n-1})\right]^T \in \mathcal{C}^n, \tag{10}$$

where

$$T_i(\mathcal{N}_i) = \mathrm{Re}(T_i(\mathcal{N}_i)) + \mathrm{i}\,\mathrm{Im}(T_i(\mathcal{N}_i)) \in \mathcal{C}, \tag{11}$$

is a complex variable, which quantifies the position and momentum quadrature transmission (i.e., gain) of the $i$-th Gaussian sub-channel $\mathcal{N}_i$, in the phase space $\mathcal{S}$, with real and imaginary parts

$$0 \leq \mathrm{Re}\,T_i(\mathcal{N}_i) \leq 1/\sqrt{2}, \text{ and } 0 \leq \mathrm{Im}\,T_i(\mathcal{N}_i) \leq 1/\sqrt{2}. \tag{12}$$

Particularly, the $T_i(\mathcal{N}_i)$ variable has the squared magnitude of

$$|T_i(\mathcal{N}_i)|^2 = \mathrm{Re}\,T_i(\mathcal{N}_i)^2 + \mathrm{Im}\,T_i(\mathcal{N}_i)^2 \in \mathbb{R}, \tag{13}$$

where

$$\mathrm{Re}\,T_i(\mathcal{N}_i) = \mathrm{Im}\,T_i(\mathcal{N}_i). \tag{14}$$

The Fourier-transformed transmittance of the $i$-th sub-channel $\mathcal{N}_i$ (resulted from CVQFT operation at Bob) is denoted by

$$|F(T_i(\mathcal{N}_i))|^2. \tag{15}$$

The $n$-dimensional zero-mean, circular symmetric complex Gaussian noise vector $\Delta \in \mathcal{CN}\left(0, \sigma_\Delta^2\right)_n$, of the quantum channel $\mathcal{N}$, is evaluated as

$$\Delta = \left(\Delta_0, ..., \Delta_{n-1}\right)^T \in \mathcal{CN}\left(0, \mathbf{K}_\Delta\right), \tag{16}$$

where

$$\mathbf{K}_\Delta = \mathbb{E}\left[\Delta \Delta^\dagger\right], \tag{17}$$

with independent, zero-mean Gaussian random components

$$\Delta_{x_i} \in \mathbb{N}\left(0, \sigma_{\mathcal{N}_i}^2\right), \text{ and } \Delta_{p_i} \in \mathbb{N}\left(0, \sigma_{\mathcal{N}_i}^2\right), \tag{18}$$

with variance $\sigma_{\mathcal{N}_i}^2$, for each $\Delta_i$ of a Gaussian sub-channel $\mathcal{N}_i$, which identifies the Gaussian noise of the $i$-th sub-channel $\mathcal{N}_i$ on the quadrature components $x_i, p_i$ in the phase space $\mathcal{S}$. Thus $F(\Delta) \in \mathcal{CN}\left(0, \sigma_{\Delta_i}^2\right)$, where

$$\sigma_{\Delta_i}^2 = 2\sigma_{\mathcal{N}_i}^2. \tag{19}$$

The CVQFT-transformed noise vector can be rewritten as

$$F(\Delta) = \left(F(\Delta_0), ..., F(\Delta_{n-1})\right)^T, \tag{20}$$

with independent components $F\left(\Delta_{x_i}\right) \in \mathbb{N}\left(0, \sigma_{\mathcal{N}_i}^2\right)$ and $F\left(\Delta_{p_i}\right) \in \mathbb{N}\left(0, \sigma_{\mathcal{N}_i}^2\right)$ on the quadratures, for each $F(\Delta_i)$.

### 2.1.1 Multiuser Quadrature Allocation (MQA)

In a MQA multiple access multicarrier CVQKD, a given user $U_k, k = 0, ..., K-1$, where $K$ is the number of total users, is characterized via $m$ subcarriers, formulating an $\mathcal{M}_{U_k}$ logical channel of $U_k$,

$$\mathcal{M}_{U_k} = \left[\mathcal{N}_{U_k, 0}, ..., \mathcal{N}_{U_k, m-1}\right]^T, \tag{21}$$

where $\mathcal{N}_{U_k, i}$ is the $i$-th sub-channel of $\mathcal{M}_{U_k}$. For a detailed description of MQA for multicarrier CVQKD see [3].

The general model of AMQD-MQA is depicted in Fig. 1 [3], [5].

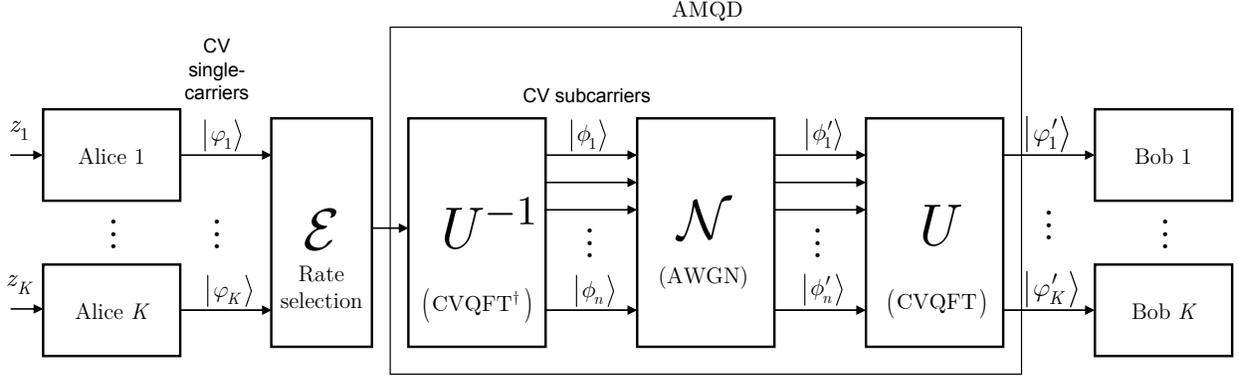

**Figure 1.** The AMQD-MQA multiple access scheme with multiple independent transmitters and multiple receivers [3]. The modulated Gaussian CV single-carriers are transformed by a unitary operation (inverse CVQFT) at the $\mathcal{E}$ encoder, which outputs the $n$ Gaussian subcarrier CVs. The parties send the $|\varphi_k\rangle$ single-carrier Gaussian CVs with variance $\sigma^2_{\omega_{0,k}}$ to Alice. In the rate-selection phase, the encoder determines the transmit users. The data states of the transmit users are then fed into the CVQFT$^\dagger$ operation. The $|\phi_i\rangle$ Gaussian subcarrier CVs have a variance $\sigma^2_\omega$ per quadrature components. The Gaussian CVs are decoded by the CVQFT unitary operation. Each $|\varphi'_k\rangle$ is received by Bob $k$ (AWGN – additive white Gaussian noise).

## 2.2 Private Classical Rate Curves

The secret key adaption method utilizes $r+2$ rate curves for the sub-channels, defined via set

$$\mathcal{S} = \{R_{\min}(\mathcal{N}_i), R(q), R_{\max}(\mathcal{N}_i)\}, q = 0\ldots r-1. \tag{22}$$

Specifically, for all sub-channels $\mathcal{N}_i$, $i = 0,\ldots,l-1$ a given rate curve is selected from $\mathcal{S}$ (22) according to the sub-channel conditions. In particular, a $R(q)$ rate curve refers to the transmission rate of *private classical information* (private rate) over $\mathcal{N}_i$, with the relation

$$R_{\min}(\mathcal{N}_i) < R(0)\ldots < R(r-1) < R_{\max}(\mathcal{N}_i), \tag{23}$$

where $R(q)$ is referred to as the *target* private rate at an $R(q-1)$ *actual* private rate in an iteration procedure.

Without loss of generality, assuming a reverse reconciliation, a given target private rate $R(q)$ is defined as

$$\begin{aligned} R(q) &\leq P(\mathcal{N}_i) \\ &= \lim_{n\to\infty} \frac{1}{n} \max_{\forall p_i, \rho_i} (\chi_{AB}(\mathcal{N}_i) - \chi_{BE}(\mathcal{N}_i)), \end{aligned} \tag{24}$$

where $P(\mathcal{N}_i)$ is the private classical capacity of $\mathcal{N}_i$, $\chi_{AB}(\mathcal{N}_i)$ and $\chi_{BE}(\mathcal{N}_i)$ are the Holevo information of Alice (transmitter) and Bob (receiver), and Bob and Eve (eavesdropper), respectively.

# 3 Secret Key Rate Adapting with Minimized Error Rate

**Theorem 1.** A *given target secret key rate* $S^*(\mathcal{N}) = \sum_{i=0}^{l-1} R_i(\mathcal{N}_i)$, *where* $R_i(\mathcal{N}_i)$ *is the private rate of sub-channel* $\mathcal{N}_i$, *can be achieved over the l sub-channels such that the error rate of all sub-channels is minimized.*

*Proof.*
The proof focuses on a single sub-channel $\mathcal{N}_i$ for the private transmission of a single quadrature component $x_i$ (or $p_i$), which refers to a $x_i \in \mathbb{N}(0, \sigma_\omega^2)$ position or a $p_i \in \mathbb{N}(0, \sigma_\omega^2)$ momentum quadrature of the $i$-th subcarrier, respectively.
Let
$$\nu_i(\mathcal{N}_i) = \sigma_{\mathcal{N}_i}^2 \Big/ \left| F(T_i(\mathcal{N}_i)) \right|^2 \tag{25}$$
of the $i$-th, $i = 0,\ldots, l-1$ sub-channel $\mathcal{N}_i$.

Let $R_{\min}(\mathcal{N}_i)$ and $R_{\max}(\mathcal{N}_i)$ be the minimal and maximal private classical information transmission rates selected for $\mathcal{N}_i$. The private rates are referring to the transmission of a given quadrature $x_i$.

These rate curves, $R(q) > R(q-1)$, allow us to reach a target secret key rate $S^*(\mathcal{N})$ over the sub-channels with a minimized error rate in a multicarrier CVQKD setting. Specifically, it requires a rigorously defined iterative condition on the selection of the sub-channels for each target rate $R(q)$.

From $R_{\min}(\mathcal{N}_i)$ and $R_{\max}(\mathcal{N}_i)$, an *adaption region* A can be characterized with $r$ rate curves inside the region. Define $r$ rate curves for the transmission of $x_i$ in the region of A
$$\mathrm{A} = \left[ R_{\min}(\mathcal{N}_i), R_{\max}(\mathcal{N}_i) \right], \tag{26}$$
as
$$R(q),\ q = 0\ldots r-1, \tag{27}$$
such that (23) holds.

Precisely, at a given private rate $R(\mathcal{N}_i)$, (25) is referred to as $\nu_i(R(\mathcal{N}_i))$,
$$\nu_i(R(\mathcal{N}_i)) = \sigma_{R(\mathcal{N}_i)}^2 \Big/ \left| F(T(R(\mathcal{N}_i))) \right|^2, \tag{28}$$
where $\sigma_{R(\mathcal{N}_i)}^2$ is the noise variance of $\mathcal{N}_i$ at $R(\mathcal{N}_i)$, while $F(T(R(\mathcal{N}_i)))$ is the transmittance coefficient of $\mathcal{N}_i$ at $R(\mathcal{N}_i)$. Note that in function of (28), after a scaling the rate curves of (27) are almost parallel to $R_{\min}(\mathcal{N}_i)$. By theory, at $R(\mathcal{N}_i) = 0$ (28) is directly defined from $\sigma_{\mathcal{N}_i}^2$ is the noise variance and $T_i(\mathcal{N}_i)$ is the transmittance coefficient of $\mathcal{N}_i$, as given in (25).
In function of $\nu_i(R(\mathcal{N}_i))$, the SNR of $\mathcal{N}_i$ at a given $R(\mathcal{N}_i)$ is expressed as [17]

$$\text{SNR}\left(R\left(\mathcal{N}_i\right)\right) = 10\log_{10}\tfrac{1}{\nu_i(R(\mathcal{N}_i))}. \tag{29}$$

To step forward, we have to focus on the behavior of parameter $\nu_i(\cdot)$ at an increased transmission rate.

Particularly, let $\underline{R}(\mathcal{N}_i)$ be the current private rate and $R(\mathcal{N}_i)$ be the target private rate for a sub-channel $\mathcal{N}_i$, such that $\underline{R}(\mathcal{N}_i) < R(\mathcal{N}_i) < \widehat{R}(\mathcal{N}_i)$. Then let $\delta_{\nu_i}(R(\mathcal{N}_i))$ identify the cumulative $\nu_i(R(\mathcal{N}_i))$ parameter of $\mathcal{N}_i$ at an increased (target) private rate $R(\mathcal{N}_i)$, evaluated via the following iteration:

$$\delta_{\nu_i}\left(R\left(\mathcal{N}_i\right)\right) = \delta_{\nu_i}\left(\underline{R}\left(\mathcal{N}_i\right)\right) + \Delta_{\nu_i}\left(R\left(\mathcal{N}_i\right), \widehat{R}\left(\mathcal{N}_i\right)\right), \tag{30}$$

where $\Delta_{\nu_i}$ identifies the difference of $\nu_i$ at $R(\mathcal{N}_i)$ and $\widehat{R}(\mathcal{N}_i)$ as

$$\Delta_{\nu_i} = \nu_i\left(R\left(\mathcal{N}_i\right)\right) - \nu_i\left(\widehat{R}\left(\mathcal{N}_i\right)\right), \tag{31}$$

the iteration (30) at no transmission, $R(\mathcal{N}_i) = 0$, identifies $\nu_i(\mathcal{N}_i)$ as

$$\delta_{\nu_i}\left(R\left(\mathcal{N}_i\right) = 0\right) = \nu_i\left(\mathcal{N}_i\right), \tag{32}$$

while for any $R(\mathcal{N}_i) > \underline{R}(\mathcal{N}_i) > 0$,

$$\nu_i\left(R\left(\mathcal{N}_i\right)\right) < \nu_i\left(\underline{R}\left(\mathcal{N}_i\right)\right). \tag{33}$$

To conclude, from (30) follows that a rate increment from $\underline{R}(\mathcal{N}_i)$ to $R(\mathcal{N}_i)$ also increases $\delta_{\nu_i}(\underline{R}(\mathcal{N}_i))$ by $\Delta_{\nu_i}(R(\mathcal{N}_i), \widehat{R}(\mathcal{N}_i))$, thus for any $R(\mathcal{N}_i) > \underline{R}(\mathcal{N}_i)$, the following relation holds [17]:

$$\delta_{\nu_i}\left(R\left(\mathcal{N}_i\right)\right) > \delta_{\nu_i}\left(\underline{R}\left(\mathcal{N}_i\right)\right). \tag{34}$$

Let us then define $r$ private transmission curves in the adaption region A of $\mathcal{N}_i$. The aim of the iterative secret key adaption scheme is to provide a rate increment in each step by selecting that sub-channel $\mathcal{N}_i$, for which (30) is minimal. Specifically, it is a convenient approach because this sub-channel provides the best condition for the transmission.

As we show, by using this sub-channel, the increased rate $\widehat{R}$ can be achieved with a minimized error rate, but at the same time, it keeps the target secret key rate. Therefore, applying the sub-channel selection procedure with respect to the iterative condition of (30), a desired target secret key rate can be achieved such that the transmission is adapted to not just the sub-channel conditions, but also to yield a minimized error rate for all sub-channels.

Applying (30) for a $R_{\min}(\mathcal{N}_i)$ private rate is as follows. An $R_{\min}(\mathcal{N}_i)$ target private rate over $\mathcal{N}_i$, $\delta_{\nu_i}(R_{\min}(\mathcal{N}_i))$ is yielded via $\nu_i(\mathcal{N}_i)$ and $\Delta_{\nu_i}(R_{\min}(\mathcal{N}_i), R(0))$ derived from $R(0) \in A$ as

$$\delta_{\nu_i}\left(R_{\min}\left(\mathcal{N}_i\right)\right) = \nu_i\left(\mathcal{N}_i\right) + \Delta_{\nu_i}\left(R_{\min}\left(\mathcal{N}_i\right), R\left(0\right)\right). \tag{35}$$

Applying (30) to the $r$ rate curves $R(q) \in A$, $q = 0...r-2$ at a target rate $R(q)$ results in $\delta_{\nu_i}(R(q))$ as

$$\delta_{\nu_i}(R(q)) = \delta_{\nu_i}(R(q-1)) + \Delta_{\nu_i}(R(q), R(q+1)), \tag{36}$$

Specifically, in each step for a given target $R(q)$, the method selects that $\mathcal{N}_i$, for which (36) is minimal, because that sub-channel provides the best conditions. Thus, the secret key adaption is an iterative process and depends on the $\delta_{\nu_i}(R(q-1))$ parameter obtained at $R(q-1)$ and on $\Delta_{\nu_i}(R(q), R(q+1))$.

In particular, for $q = 0$, (36) yields
$$\begin{aligned}\delta_{\nu_i}(R(0)) &= \delta_{\nu_i}(R_{\min}(\mathcal{N}_i)) + \Delta_{\nu_i}(R(0), R(1)) \\ &= \nu_i(\mathcal{N}_i) + \Delta_{\nu_i}(R_{\min}(\mathcal{N}_i), R(0)) + \Delta_{\nu_i}(R(0), R(1)),\end{aligned} \tag{37}$$

where $R(1) \in A$. While, for $q = r-1$, (30) results in
$$\begin{aligned}\delta_{\nu_i}(R(r-1)) &= \delta_{\nu_i}(R(r-2)) + \Delta_{\nu_i}(R(r-1), R_{\max}(\mathcal{N}_i)) \\ &= \left[\nu_i(\mathcal{N}_i) + \Delta_{\nu_i}(R_{\min}(\mathcal{N}_i), R(0)) + \sum_{k=0}^{r-2}\Delta_{\nu_i}(R(k), R(k+1))\right] \\ &\quad + \Delta_{\nu_i}(R(r-1), R_{\max}(\mathcal{N}_i)),\end{aligned} \tag{38}$$

where $R(k) \in A$, $k = 0...r-2$. For $R_{\max}(\mathcal{N}_i)$, by definition $\delta_{\nu_i}(R_{\max}(\mathcal{N}_i)) = +\infty$ [17].

The distribution of a sample set $\delta_{\nu_i}(R(q))$ for $m$ sub-channels, $i = 0,...,m-1$, in a low-SNR CVQKD scenario is illustrated in Fig. 2(a). The SNR in Fig. 2(b) is derived from the $\delta_{\nu_i}(R(q))$ set of the $m$ sub-channels as $\mathrm{SNR}(\nu_i(R(q))) = 10\log_{10}(1/(\nu_i(R(q))))$.

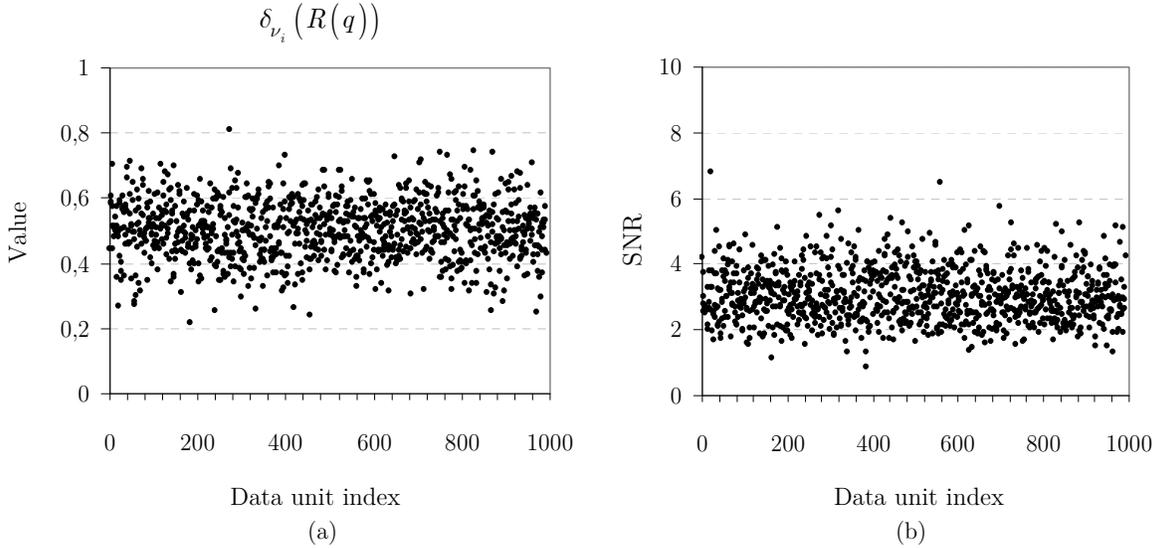

**Figure 2.** The distribution of a low-SNR set $\delta_{\nu_i}(R(q))$, $i = 0,...,m-1$, for $m = 1000$ sub-channels **(a)**. The SNR is derived from $\delta_{\nu_i}(R(q))$ for all sub-channels **(b)**.

Next, we show that for an arbitrary target private rate $R(\mathcal{N}_i) > 0$, the iterative condition on $\delta_{\nu_i}(R(\mathcal{N}_i))$ (see (36)) provides a minimized error rate over the selected $\mathcal{N}_i$.

Let $\mathcal{B}$ be the bit error rate, and let $\mathcal{B}\left[R(q)_{\delta_{\nu_i}(R(q-1))}\right]$ refer to the bit error rate of $\mathcal{N}_i$ at target private rate $R(q)$, at an actual rate $R(q-1)$, and $\delta_{\nu_i}(R(q-1))$. Then, at a given $R(q)$, selecting that $\mathcal{N}_i$, $i=0,\ldots,l-1$ from the total $l$, for which $\delta_{\nu_i}(R(q-1))$ is minimal, as

$$\xi(R(q-1)) = \min_{\forall i} \delta_{\nu_i}(R(q-1)) \tag{39}$$

yields a minimized bit error rate at a given $R(q)$ over the selected $\mathcal{N}_i$ as

$$\mathcal{B}\left[R(q)_{\xi(R(q-1))}\right] = \mathcal{B}\left[R_{\min}(\mathcal{N}_i)_{\delta_{\nu_i}(R(q-1))}\right], \tag{40}$$

where $\mathcal{B}\left[R_{\min}(\mathcal{N}_i)_{\delta_{\nu_i}(R(q-1))}\right]$ is defined as

$$\mathcal{B}\left[R_{\min}(\mathcal{N}_i)_{\nu_i(\mathcal{N}_i)+\left[\Delta_{\nu_i}(R_{\min}(\mathcal{N}_i),R(0))+\sum_{k=0}^{q-1}\Delta_{\nu_i}(R(k),R(k+1))\right]}\right]$$
$$= \tfrac{1}{2}\mathrm{erfc}\mathcal{F}\left(\nu_i(\mathcal{N}_i) + \left(\Delta_{\nu_i}(R_{\min}(\mathcal{N}_i),R(0)) + \ldots + \Delta_{\nu_i}(R(q-1),R(q))\right)\right), \tag{41}$$

where function $\mathcal{F}(\cdot)$ is evaluated as

$$\mathcal{F}\left(\nu_i(\mathcal{N}_i) + \left(\Delta_{\nu_i}(R_{\min}(\mathcal{N}_i),R(0)) + \ldots + \Delta_{\nu_i}(R(q-1),R(q))\right)\right)$$
$$= \left(\sqrt{\begin{array}{c}\mathrm{SNR}(\nu_i(\mathcal{N}_i)) - \left[\mathrm{SNR}(\nu_i(R(0))) - \mathrm{SNR}(\nu_i(R_{\min}(\mathcal{N}_i)))\right] \\ -\left(\sum_{k=0}^{q-1}\left[\mathrm{SNR}(\nu_i(R(k+1))) - \mathrm{SNR}(\nu_i(R(k)))\right]\right)\end{array}}\right), \tag{42}$$

where $\mathrm{SNR}(\nu_i(R(x))) = 10\log_{10}(1/(\nu_i(R(x))))$, while $\mathrm{erfc}(\cdot)$ is the complementary error function

$$\mathrm{erfc}(x) = \tfrac{2}{\sqrt{\pi}} \int_x^\infty e^{-t^2} dt. \tag{43}$$

The aim of the error minimization procedure is to achieve (40) for all $R(\cdot)$ via the selection of that $\mathcal{N}_i$ for which $\delta_{\nu_i}(R(q-1))$ is minimal.

Let $R(k+1)_{\delta_{\nu_i}(R(k))}$ refer to a private rate $R(k+1)$ over at $\delta_{\nu_i}(R(k))$ with respect to the transmission of a single quadrature component $x_i$ (or $p_i$).

First, we apply (39) to the minimal rate $R_{\min}(\mathcal{N}_i)$ by the selection of that $\mathcal{N}_i$, for which $\nu_i(\mathcal{N}_i)$ is minimal, thus

$$\xi(0) = \min_{\forall i} \delta_{\nu_i}(0) = \min_{\forall i} \nu_i, \tag{44}$$

which yields

$$\mathcal{B}\left[R_{\min}(\mathcal{N}_i)_{\xi(0)}\right] = \mathcal{B}\left[R_{\min}(\mathcal{N}_i)_{\nu_i(\mathcal{N}_i)}\right]. \tag{45}$$

Similarly, at $R(0)$, that sub-channel is selected for the $R(0)$ rate transmission, for which $\delta_{\nu_i}(R_{\min}(\mathcal{N}_i))$ is minimal.

In particular, due to the iterative determination of $\delta_{\nu_i}(R_{\min}(\mathcal{N}_i))$, the corresponding $\mathcal{B}(R(0)_{\xi(R_{\min}(\mathcal{N}_i))})$ is yielded as

$$\mathcal{B}\left(R(0)_{\xi(R_{\min}(\mathcal{N}_i))}\right) = \mathcal{B}\left(R_{\min}(\mathcal{N}_i)_{\nu_i(\mathcal{N}_i)+\Delta_{\nu_i}(R_{\min}(\mathcal{N}_i),R(0))}\right). \quad (46)$$

Precisely, for $R(q)$, $q = 1\ldots r-1$, therefore, in each step, that sub-channel selected for the transmission, for which $\xi(R(q-1))$ is minimal, ensures that the resulting $\mathcal{B}\left(R(q)_{\xi(R(q-1))}\right)$ is evaluated as

$$\mathcal{B}\left(R(q)_{\xi(R(q-1))}\right) = \mathcal{B}\left(R_{\min}(\mathcal{N}_i)_{\delta_{\nu_i}(R(q-2))+\Delta_{\nu_i}(R(q-1),R(q))}\right). \quad (47)$$

Putting the pieces together, the utilization of (47) for $q = 1$ is as

$$\mathcal{B}\left(R(1)_{\xi(R(0))}\right) = \mathcal{B}\left(R_{\min}(\mathcal{N}_i)_{\delta_{\nu_i}(R_{\min}(\mathcal{N}_i))+\Delta_{\nu_i}(R(0),R(1))}\right), \quad (48)$$

while for $q = r-1$,

$$\mathcal{B}\left(R(r-1)_{\xi(R(r-2))}\right) = \mathcal{B}\left(R_{\min}(\mathcal{N}_i)_{\delta_{\nu_i}(R(r-3))+\Delta_{\nu_i}(R(r-2),R(r-1))}\right), \quad (49)$$

and finally, for $R_{\max}(\mathcal{N}_i)_{\xi(R(r-1))}$, the corresponding error rate is

$$\mathcal{B}\left(R_{\max}(\mathcal{N}_i)_{\xi(R(r-1))}\right) = \mathcal{B}\left(R_{\min}(\mathcal{N}_i)_{\delta_{\nu_i}(R(r-2))+\Delta_{\nu_i}(R(r-1),R_{\max}(\mathcal{N}_i))}\right). \quad (50)$$

The $\mathcal{B}$ bit error rates at private rates, $R_{\min}(\mathcal{N}_i), R(0), \ldots, R_{\min}(r-1), R_{\max}(\mathcal{N}_i)$, for a given sub-channel $\mathcal{N}_i$ in function of $\nu_i$ (low-SNR scenario) are summarized in Fig. 3 for the range $\nu_i = [0.1, 0.3]$ (a), and $\nu_i = [0.3, 0.9]$ (b).

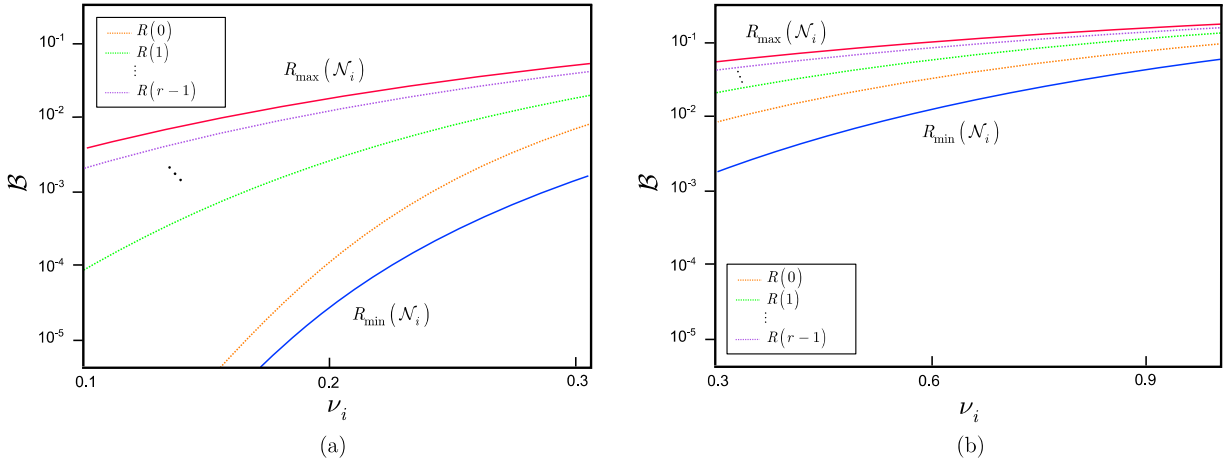

**Figure 3.** The $\mathcal{B}$ bit error rates at $R_{\min}(\mathcal{N}_i), R(0), \ldots, R(r-1), R_{\max}(\mathcal{N}_i)$ in a low-SNR scenario for a given sub-channel $\mathcal{N}_i$ at $\nu_i = [0.1, 0.3]$ **(a)** and $\nu_i = [0.3, 0.9]$ **(b)**.

The iterative secret key adapting method with the error minimization is summarized as follows.

**Algorithm (Iterative Secret Key Adapting)**

1. Let $R_{\min}(\mathcal{N}_i), R_{\max}(\mathcal{N}_i)$ be the minimal and maximal private transmission rates selected for all sub-channels $\mathcal{N}_i$, $i = 0,\ldots,l-1$.

2. For a given $\mathcal{N}_i$, let $\nu_i(\mathcal{N}_i) = \sigma^2_{\mathcal{N}_i} \big/ |F(T(\mathcal{N}_i))|^2$ at noise variance $\sigma^2_{\mathcal{N}_i}$ and transmission coefficient $F(T_i(\mathcal{N}_i))$. Define $r$ curves $R(q)$, $q = 0\ldots r-1$ in the adaption region $\mathrm{A} = [R_{\min}(\mathcal{N}_i), R_{\max}(\mathcal{N}_i)]$, with relation $R_{\min}(\mathcal{N}_i) < R(0)\ldots < R(r-1) < R_{\max}(\mathcal{N}_i)$.

3. For target private rate $R_{\min}(\mathcal{N}_i)$, determine $\delta_{\nu_i}(R_{\min}(\mathcal{N}_i))$ as $\delta_{\nu_i}(R_{\min}(\mathcal{N}_i)) = \nu_i(\mathcal{N}_i) + \Delta_{\nu_i}(R_{\min}(\mathcal{N}_i), R(0))$. For the $R(q)$ $q = 0\ldots r-2$ curves of the adaption region $\mathrm{A}$, compute $\delta_{\nu_i}(R(q))$ as $\delta_{\nu_i}(R(q)) = \delta_{\nu_i}(R(q-1)) + \Delta_{\nu_i}(R(q), R(q+1))$. At $q = r-1$, use $\delta_{\nu_i}(R(r-1)) = \delta_{\nu_i}(R(r-2)) + \Delta_{\nu_i}(R(r-1), R_{\max}(\mathcal{N}_i))$. At $R_{\max}(\mathcal{N}_i)$, use $\delta_{\nu_i}(R_{\max}(\mathcal{N}_i)) = +\infty$.

4. Utilize the adaption method: At a target rate $R(q)$, $q = 0\ldots r-1$, select that sub-channel $\mathcal{N}_i$ for which $\delta_{\nu_i}(R(q-1))$ is minimal. At $R_{\min}(\mathcal{N}_i)$, select that $\mathcal{N}_i$ for which $\nu_i(\mathcal{N}_i)$ is minimal. At $R(0)$ select $\mathcal{N}_i$ for which $\delta_{\nu_i}(R_{\min}(\mathcal{N}_i))$ is minimal, while at $R_{\max}(\mathcal{N}_i)$, select $\mathcal{N}_i$ for which $\delta_{\nu_i}(R(r-1))$ is minimal.

5. Repeat the steps until $S^*(\mathcal{N}) \leq \sum_{i=0}^{l-1} R_i(\mathcal{N}_i)$, where $S^*(\mathcal{N})$ is the desired secret key rate over the $l$ sub-channels, and $R_i(\mathcal{N}_i)$ is the private transmission rate of $\mathcal{N}_i$ at a minimized $\mathcal{B}$ for all $\mathcal{N}_i$.

∎

The resulting $\mathcal{B}$ bit error rates of the secret key rate adapting method for a given sub-channel $\mathcal{N}_i$ in function of $\nu_i$ are summarized in Fig. 4. Parameter $\nu_i$ is scaled for the SNR with 5 dB steps in the range of $[15, -5]$ as $\nu_i(R_i(x)) = 10^{-\mathrm{SNR}(\nu_i(R(x)))/10}$

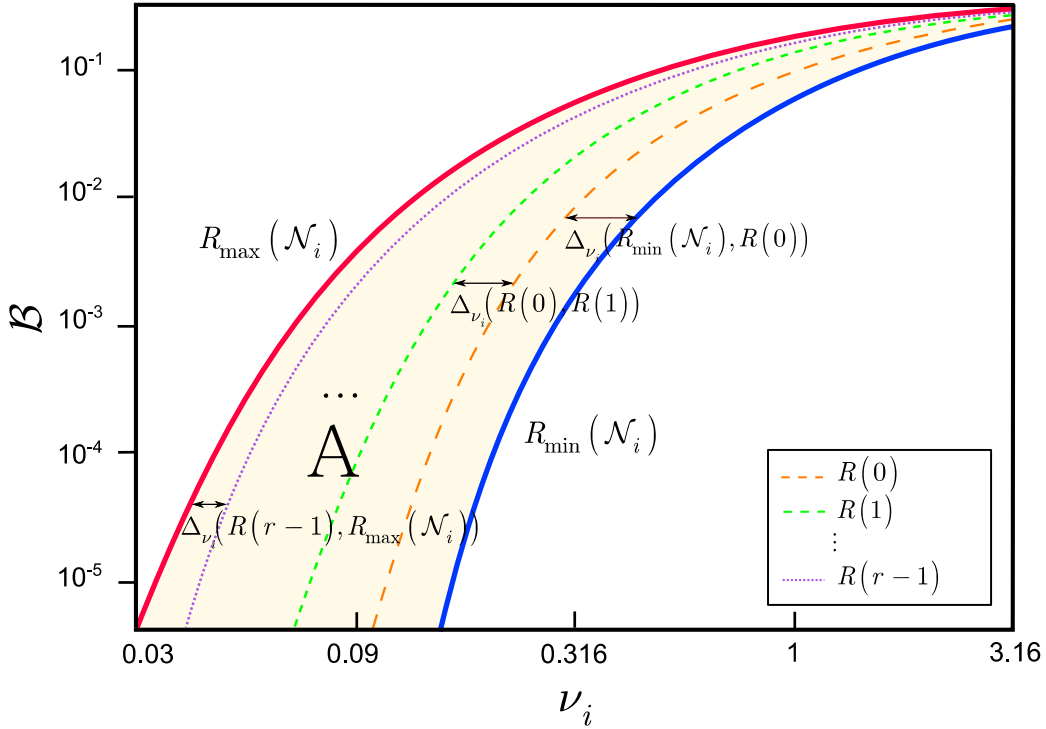

**Figure 4.** The $\mathcal{B}$ bit error rate of secret key adapting for a sub-channel $\mathcal{N}_i$ in function of $\nu_i\left(R_i\left(x\right)\right) = 10^{-\text{SNR}\left(\nu_i\left(R(x)\right)\right)/10}$. The adaption is made via $r$ rate curves $R(q)$, $q = 0...r-1$, $R_{\min}\left(\mathcal{N}_i\right) < R(0) ... < R(r-1) < R_{\max}\left(\mathcal{N}_i\right)$ in the $\text{A} = \left[R_{\min}\left(\mathcal{N}_i\right), R_{\max}\left(\mathcal{N}_i\right)\right]$ adaption region (shaded area).

## 4 Multiuser Multicarrier CVQKD Scenario

This section extends the results for a multiuser multicarrier CVQKD scenario.

**Lemma 1.** *The secret key adaption can be extended to a $U_k, k = 0,...,K-1$ multiuser setting, where $K$ is the number of users, to achieve target secret key rate $S^*\left(\mathcal{M}_{U_k}\right)$ with minimized error rate over the $m$ sub-channels of $\mathcal{M}_{U_k}$ of $U_k$, for $\forall k$.*

*Proof.*
The proof focuses on a given logical channel $\mathcal{M}_{U_k} = \left[\mathcal{N}_{U_k,0},...,\mathcal{N}_{U_k,m-1}\right]^T$ of a user $U_k$, where $\mathcal{N}_{U_k,i}$, $i = 0,...,m-1$ is the $i$-th sub-channel.

Let $K$ be the number of transmit users, and select a given $U_k, k = 0,...,K-1$. Let $S^*\left(\mathcal{M}_{U_k}\right)$ be the target secret key rate of users over $\mathcal{M}_{U_k}$, and let $R\left(\mathcal{N}_{U_k,i}\right)$ be the private rate of $\mathcal{N}_{U_k,i}$. The steps of the extension are summarized as follows.

Let
$$\nu_i\left(\mathcal{N}_{U_k,i}\right) = \sigma^2_{\mathcal{N}_{U_k,i}} \bigg/ \left|F\left(T\left(\mathcal{N}_{U_k,i}\right)\right)\right|^2, \tag{51}$$

where $\sigma^2_{\mathcal{N}_{U_k,i}}$ is the noise variance of $\mathcal{N}_{U_k,i}$, while $T_i\left(\mathcal{N}_{U_k,i}\right)$ is the transmittance coefficient of $\mathcal{N}_{U_k,i}$. At a given rate curve $R\left(\mathcal{N}_{U_k,i}\right) > 0$,

$$\nu_i\left(R\left(\mathcal{N}_{U_k,i}\right)\right) = \sigma^2_{R\left(\mathcal{N}_{U_k,i}\right)} \bigg/ \left|F\left(T\left(R\left(\mathcal{N}_{U_k,i}\right)\right)\right)\right|^2, \tag{52}$$

where $\sigma^2_{R\left(\mathcal{N}_{U_k,i}\right)}$ is the noise variance at $R\left(\mathcal{N}_{U_k,i}\right)$, while $T_i\left(R\left(\mathcal{N}_{U_k,i}\right)\right)$ is the transmittance coefficient of $\mathcal{N}_{U_k,i}$ at $R\left(\mathcal{N}_{U_k,i}\right)$, respectively. Apply the secret key rate adaption method over the set $\mathcal{M}_{U_k}$ of $m$ $\mathcal{N}_{U_k,i}$, sub-channels of $\mathcal{M}_{U_k}$, until

$$S^*\left(\mathcal{M}_{U_k}\right) \leq \sum_{i=0}^{m-1} R\left(\mathcal{N}_{U_k,i}\right). \tag{53}$$

Apply the steps for all transmit users $U_k$ for their $\mathcal{M}_{U_k}$ sets of $m$ sub-channels. Therefore, one can utilize (47) at a given $R\left(\mathcal{N}_{U_k,i}\right) = R_i(q)$, where $R_i(q)$ refers to the $R(q)$ curve of $\mathcal{N}_{U_k,i}$, and from (32) follows $\delta_{\nu_i}\left(R\left(\mathcal{N}_{U_k,i}\right) = 0\right) = \nu_i\left(\mathcal{N}_{U_k,i}\right)$, which yields the error rate for $\mathcal{N}_{U_k,i}$ of $\mathcal{M}_{U_k}$ as

$$\mathcal{B}\left(R\left(\mathcal{N}_{U_k,i}\right)_{\xi(R(q-1))}\right) = \mathcal{B}\left(R_{\min}\left(\mathcal{N}_{U_k,i}\right)_{\nu_i\left(\mathcal{N}_{U_k,i}\right) + \left(\Delta_{\nu_i}\left(R_{\min}\left(\mathcal{N}_{U_k,i}\right), R_i(0)\right) + \sum_{k=0}^{q-1}\Delta_{\nu_i}\left(R_i(k), R_i(k+1)\right)\right)}\right). \tag{54}$$

∎

## 4.1 Variance Adaption for an Equalized Error Rate for Users

In this section, we propose a modulation variance adaption method to achieve an equally minimized error rate for the sub-channels of a given user. The results can be extended to an arbitrary number of users.

**Theorem 2.** *For all $U_k$, $k = 0,...,K-1$, the error rate of the $\mathcal{N}_{U_k,i}$, $i = 0,...,m-1$ sub-channels of $\mathcal{M}_{U_k}$ of user $U_k$ can be equally minimized via the a $\tilde{\sigma}^2_\omega = \sigma^2_\omega + \Delta_{\sigma^2_{\omega_i}}$ modulation variance correction, where $\Delta_{\sigma^2_{\omega_i}} > 0$.*

*Proof.*
Let
$$R\left(\mathcal{N}_{U_k,i}\right) = R_i(q-1), q = 0,...,r-1, \tag{55}$$

and $R(\mathcal{N}_{U_k,i})$ be the target private rate,

$$R(\mathcal{N}_{U_k,i}) = R_i(q), \tag{56}$$

where $R_i(q)$ refers to the $R(q)$ curve of $\mathcal{N}_{U_k,i}$, with relation to $R_i(q-1) < R_i(q)$, and let us identify $R_i(-1) = R_{\min}(\mathcal{N}_{U_k,i})$ and $R_i(r) = R_{\max}(\mathcal{N}_{U_k,i})$.

Precisely, for a given $U_k$ at $R(\mathcal{N}_{U_k,i})$, the minimal $\delta_{\nu_i}(R(\mathcal{N}_{U_k,i}))$ parameter for the $\mathcal{N}_{U_k,i}$, $i = 0, \ldots, m-1$ sub-channels of the set $\mathcal{M}_{U_k}$ is evaluated as

$$\xi_{U_k} = \min_{\forall i \in \mathcal{M}_{U_k}} \delta_{\nu_i}(\underline{R}(\mathcal{N}_{U_k,i})). \tag{57}$$

Let $\sigma_\omega^2$ refer to the input modulation variance of the $i$-th subcarrier of $U_k$. Specifically, using the expression of $\xi_{U_k}$ in (57), the $\sigma_\omega^2$ modulation variance of the $i$-th subcarrier is corrected by $\Delta_{\sigma_{\omega_i}^2}$ as

$$\Delta_{\sigma_{\omega_i}^2} = \delta_{\nu_i}(\underline{R}(\mathcal{N}_{U_k,i})) - \xi_{U_k}, \tag{58}$$

yielding a modulation variance increment

$$\tilde{\sigma}_\omega^2 = \sigma_\omega^2 + \Delta_{\sigma_{\omega_i}^2} \tag{59}$$

for the input of $\mathcal{N}_{U_k,i}$.

In particular, using (59), the $\varphi_{\delta_{\nu_i}}(R(\mathcal{N}_{U_k,i}))$ resulting $\delta_{\nu_i}$ parameter for $\mathcal{N}_{U_k,i}$ at a target rate $R(\mathcal{N}_{U_k,i})$ is therefore

$$\varphi_{\delta_{\nu_i}}(R(\mathcal{N}_{U_k,i})) = \delta_{\nu_i}(R(\mathcal{N}_{U_k,i})) - \xi_{U_k}, \tag{60}$$

from which the $\Delta_{\mathrm{SNR}}(\varphi_{\delta_{\nu_i}}(R(\mathcal{N}_{U_k,i})))$ SNR increment of $\mathcal{N}_{U_k,i}$ at a given $R(\mathcal{N}_{U_k,i}) = R(q)$ is as

$$\begin{aligned}&\Delta_{\mathrm{SNR}}(\varphi_{\delta_{\nu_i}}(R(\mathcal{N}_{U_k,i})))\\&= 10\log_{10}\tfrac{1}{\xi_{U_k}}\\&\quad -\mathcal{F}\big(\nu_i(\mathcal{N}_{U_k,i}) + (\Delta_{\nu_i}(R_{\min}(\mathcal{N}_{U_k,i}), R(0)) + \ldots + \Delta_{\nu_i}(R(q), R(q+1)))\big),\end{aligned} \tag{61}$$

where $\mathcal{F}(\cdot)$ is specified in (42).

Therefore, the $\mathcal{B}\big(R(\mathcal{N}_{U_k,i})_{\xi_{U_k}}\big)$ error rate for all $\mathcal{N}_{U_k,i}$ at an arbitrary $R(\mathcal{N}_{U_k,i})_{\delta_{\nu_i}(\underline{R}(\mathcal{N}_{U_k,i}))}$ is equally minimized by $\xi_{U_k}$ via (41) as

$$\mathcal{B}\left(R\left(\mathcal{N}_{U_k,i}\right)_{\xi_{U_k}}\right)$$
$$= \min_{\forall i \in \mathcal{M}_{U_k}} \mathcal{B}\left(R\left(\mathcal{N}_{U_k,i}\right)_{\delta_{\nu_i}\left(\underline{R}\left(\mathcal{N}_{U_k,i}\right)\right)}\right) \quad (62)$$
$$= \mathcal{B}\left(R_{\min}\left(\mathcal{N}_{U_k,i}\right)_{\delta_{\nu_i}\left(\underline{R}\left(\mathcal{N}_{U_k,i}\right)\right)}\right), i = 0,\ldots,m-1.$$

The formula of (62) proves the minimal error rate at arbitrary $S^*\left(\mathcal{M}_{U_k}\right)$ over $\mathcal{M}_{U_k}$ for all $\mathcal{N}_{U_k,i}$ sub-channels of $U_k$.

Without loss of generality, the results can be extended for all $K$ users to achieve minimized equalized error rate over all $\mathcal{M}_{U_k}$, $k = 0,\ldots,K-1$ logical channels.

In a single user setting, e.g., $K = 1$, the method provides an equal, minimized error rate over the $l$ sub-channels.

∎

A numerical evidence is included in the Supplemental Information.

## 5 Conclusions

We defined an iterative secret key adaption method for multicarrier CVQKD. The scheme provides a minimized error rate using the utilization of an adaptive private classical information transmission through the sub-channels. The private classical transmission is realized through pre-defined private rate curves, which characterize an adaption region for each sub-channel to find the best conditions for the transmission at a given private classical rate. The method allows us to reach a given target secret key rate with optimal transmit conditions and minimized error rate for all sub-channels. The scheme requires no complex calculations or sophisticated computational tools, allowing for easy implementation for experimental CVQKD scenarios.

## Acknowledgements

This work was partially supported by the GOP-1.1.1-11-2012-0092 (*Secure quantum key distribution between two units on optical fiber network*) project sponsored by the EU and European Structural Fund, by the Hungarian Scientific Research Fund - OTKA K-112125, and by the COST Action MP1006.

## References

[1] S. Imre and L. Gyongyosi. *Advanced Quantum Communications - An Engineering Approach.* Wiley-IEEE Press (New Jersey, USA), (2012).

[2] L. Gyongyosi, Adaptive Multicarrier Quadrature Division Modulation for Continuous-variable Quantum Key Distribution, *arXiv:1310.1608* (2013).

[3] L. Gyongyosi, Multiuser Quadrature Allocation for Continuous-Variable Quantum Key Distribution, *arXiv:1312.3614* (2013).

[4] L. Gyongyosi, Singular Layer Transmission for Continuous-Variable Quantum Key Distribution, *arXiv:1402.5110* (2014).

[5] L. Gyongyosi, Security Thresholds of Multicarrier Continuous-Variable Quantum Key Distribution, *arXiv:1404.7109* (2014).

[6] L. Gyongyosi, Multidimensional Manifold Extraction for Multicarrier Continuous-Variable Quantum Key Distribution, *arXiv:1405.6948* (2014).

[7] L. Gyongyosi, Subcarrier Domain of Multicarrier Continuous-Variable Quantum Key Distribution, *arXiv:1406.6949* (2014).

[8] L. Gyongyosi, Adaptive Quadrature Detection for Multicarrier Continuous-Variable Quantum Key Distribution, *arXiv:1408.6493* (2014).

[9] L. Gyongyosi, Distribution Statistics and Random Matrix Formalism for Multicarrier Continuous-Variable Quantum Key Distribution, *arXiv:1410.8273* (2014).

[10] L. Gyongyosi, Gaussian Quadrature Inference for Multicarrier Continuous-Variable Quantum Key Distribution, *arXiv:1504.05574* (2015).

[11] L. Gyongyosi, S. Imre, Geometrical Analysis of Physically Allowed Quantum Cloning Transformations for Quantum Cryptography, *Information Sciences,* Elsevier, pp. 1-23, DOI: 10.1016/j.ins.2014.07.010 (2014).

[12] S. Pirandola, R. Garcia-Patron, S. L. Braunstein and S. Lloyd. *Phys. Rev. Lett.* 102 050503. (2009).

[13] S. Pirandola, A. Serafini and S. Lloyd. *Phys. Rev. A* 79 052327. (2009).

[14] S. Pirandola, S. L. Braunstein and S. Lloyd. *Phys. Rev. Lett.* 101 200504 (2008).

[15] C. Weedbrook, S. Pirandola, S. Lloyd and T. Ralph. *Phys. Rev. Lett.* 105 110501 (2010).

[16] C. Weedbrook, S. Pirandola, R. Garcia-Patron, N. J. Cerf, T. Ralph, J. Shapiro, and S. Lloyd. *Rev. Mod. Phys.* 84, 621 (2012).

[17] H. Rohling, C. Fellenberg, Successive Bit Loading Concept, *OFDM Concepts for Future Communication Systems*, Springer (2011).

[18] L. Gyongyosi, Scalar Reconciliation for Gaussian Modulation of Two-Way Continuous-variable Quantum Key Distribution, *arXiv:1308.1391* (2013).

[19] P. Jouguet, S. Kunz-Jacques, A. Leverrier, P. Grangier, E. Diamanti, Experimental demonstration of long-distance continuous-variable quantum key distribution, *arXiv:1210.6216v1* (2012).

[20] M. Navascues, F. Grosshans, and A. Acin. Optimality of Gaussian Attacks in Continuous-variable Quantum Cryptography, *Phys. Rev. Lett. 97, 190502* (2006).

[21] R. Garcia-Patron and N. J. Cerf. Unconditional Optimality of Gaussian Attacks against Continuous-Variable Quantum Key Distribution. *Phys. Rev. Lett. 97, 190503* (2006).

[22] F. Grosshans, Collective attacks and unconditional security in continuous variable quantum key distribution. *Phys. Rev. Lett. 94,* 020504 (2005).

[23] M R A Adcock, P Høyer, and B C Sanders, Limitations on continuous-variable quantum algorithms with Fourier transforms, *New Journal of Physics 11 103035* (2009)

[24] S. Pirandola, S. Mancini, S. Lloyd, and S. L. Braunstein, Continuous-variable Quantum Cryptography using Two-Way Quantum Communication, *arXiv:quant-ph/0611167v3* (2008).

[25] L. Hanzo, H. Haas, S. Imre, D. O'Brien, M. Rupp, L. Gyongyosi. Wireless Myths, Realities, and Futures: From 3G/4G to Optical and Quantum Wireless, *Proceedings of the IEEE*, Volume: 100, *Issue: Special Centennial Issue*, pp. 1853-1888. (2012).

[26] D. Tse and P. Viswanath. *Fundamentals of Wireless Communication*, Cambridge University Press, (2005).

[27] D. Middlet, *An Introduction to Statistical Communication Theory: An IEEE Press Classic Reissue*, Hardcover, IEEE, ISBN-10: 0780311787, ISBN-13: 978-0780311787 (1960)

[28] S. Kay, *Fundamentals of Statistical Signal Processing, Volumes I-III*, Prentice Hall, (2013)

[29] O. S. Jahromi, *Multirate Statistical Signal Processing*, ISBN-10 1-4020-5316-9, Springer (2007).

[30] G. Heinzel, A. Rudiger, R. Schilling, Spectrum and spectral density estimation by the Discrete Fourier transform (DFT), including a comprehensive list of window functions and some new at-top windows, *http://hdl.handle.net/11858/00-001M-0000-0013-557A-5* (2002).

[31] W. H. Press, S. A. Teukolsky, W. T. Vetterling and B. P. Flannery, *Numerical Recipes in C: The Art of Scientific Computing*, ISBN : 0-521-43108-5, Cambridge University Press (1993).

[32] S. Imre, F. Balazs: *Quantum Computing and Communications – An Engineering Approach*, John Wiley and Sons Ltd, ISBN 0-470-86902-X, 283 pages (2005).

[33] D. Petz, *Quantum Information Theory and Quantum Statistics*, Springer-Verlag, Heidelberg, Hiv: 6. (2008).

[34] R. V. Meter, *Quantum Networking*, John Wiley and Sons Ltd, ISBN 1118648927, 9781118648926 (2014).

[35] L. Gyongyosi, S. Imre: Properties of the Quantum Channel, *arXiv:1208.1270* (2012).

[36] K Wang, XT Yu, SL Lu, YX Gong, Quantum wireless multihop communication based on arbitrary Bell pairs and teleportation, *Phys. Rev A*, (2014).

[37] Babar, Zunaira, Ng, Soon Xin and Hanzo, Lajos, EXIT-Chart Aided Near-Capacity Quantum Turbo Code Design. *IEEE Transactions on Vehicular Technology* (*submitted*) (2014).

[38] Botsinis, Panagiotis, Alanis, Dimitrios, Ng, Soon Xin and Hanzo, Lajos Low-Complexity Soft-Output Quantum-Assisted Multi-User Detection for Direct-Sequence Spreading and Slow Subcarrier-Hopping Aided SDMA-OFDM Systems. *IEEE Access*, PP, (99), doi:10.1109/ACCESS.2014.2322013 (2014).

[39] Botsinis, Panagiotis, Ng, Soon Xin and Hanzo, Lajos Fixed-complexity quantum-assisted multi-user detection for CDMA and SDMA. *IEEE Transactions on Communications*, vol. 62, (no. 3), pp. 990-1000, doi:10.1109/TCOMM.2014.012514.130615 (2014).

[40] L. Gyongyosi, S. Imre, Adaptive multicarrier quadrature division modulation for long-distance continuous-variable quantum key distribution, *Proc. SPIE 9123, Quantum Information and Computation XII*, 912307; doi:10.1117/12.2050095, From Conference Volume 9123, Quantum Information and Computation XII, Baltimore, Maryland, USA (2014).

[41] L. Gyongyosi: Statistical Quadrature Evolution by Inference for Continuous-Variable Quantum Key Distribution, *arXiv:1603.09247* (2016).

[42] L. Gyongyosi, S. Imre, Gaussian Quadrature Inference for Multicarrier Continuous-Variable Quantum Key Distribution, *SPIE Quantum Information and Computation XIV*, 17 - 21 Apr 2016, Baltimore, Maryland, USA (2016).

# Supplemental Information

## S.1 Numerical Evidence

This section proposes numerical evidence to demonstrate the results through a multiuser multi-carrier CVQKD environment (AMQD-MQA [3]). The numerical evidence serves demonstration purposes.

### S.1.1 Parameters

To demonstrate the results of Section 4.1, let $U_k$ be a given user with $m$ sub-channels. The parameters of the numerical evidence are summarized as follows.
The single-carrier inputs of user $U_k$,
$$x_{U_k,j} \in \mathbb{N}\left(0, \sigma^2_{\omega_0}\right), \tag{S.1}$$
have a modulation variance of $\sigma^2_{\omega_0}$ and formulate a $d$-dimensional input vector $\vec{x}_{U_k}$.
The $j$-th single-carrier is dedicated to a single-carrier channel $\mathcal{N}_{U_k,j}$. The single-carrier channel transmittance coefficient is depicted by $T\left(\mathcal{N}_{U_k,j}\right)$, $j = 0, ..., d-1$, where $d$ is the dimension of the input vector.
The single-carriers are granulated into $m$ subcarriers, where the $i$-th subcarrier is
$$x_{U_k,i} \in \mathbb{N}\left(0, \sigma^2_{\omega}\right), \tag{S.2}$$
and has a modulation variance of $\sigma^2_{\omega}$.
The $m$ sub-channels,
$$\mathcal{N}_{U_k,i}, \ i = 0, ..., m-1, \tag{S.3}$$
formulate the $\mathcal{M}_{U_k}$ logical channel of user $U_k$,
$$\mathcal{M}_{U_k} = \left[\mathcal{N}_{U_k,0}, ..., \mathcal{N}_{U_k,m-1}\right]^T. \tag{S.4}$$
The $\Delta_{x_i} \in \mathbb{N}\left(0, \sigma^2_{\mathcal{N}_{U_k,i}}\right)$ noise of $\mathcal{N}_{U_k,i}$ is added to the subcarriers, where $\sigma^2_{\mathcal{N}_{U_k,i}}$ is the noise variance of $\mathcal{N}_{U_k,i}$.
For a given sub-channel $\mathcal{N}_{U_k,i}$ of $U_k$, parameters $\nu_i\left(\mathcal{N}_{U_k,i}\right)$ and $\nu_i\left(R\left(\mathcal{N}_{U_k,i}\right)\right)$ are evaluated as
$$\nu_i\left(\mathcal{N}_{U_k,i}\right) = \sigma^2_{\mathcal{N}_{U_k,i}} \bigg/ \left|F\left(T\left(\mathcal{N}_{U_k,i}\right)\right)\right|^2, \tag{S.5}$$
and

$$\nu_i\left(R\left(\mathcal{N}_{U_k,i}\right)\right) = \sigma^2_{R\left(\mathcal{N}_{U_k,i}\right)} \Big/ \left|F\left(T\left(R\left(\mathcal{N}_{U_k,i}\right)\right)\right)\right|^2, \qquad (S.6)$$

where the $T\left(\mathcal{N}_{U_k,i}\right)$ sub-channel transmittance coefficients are estimated in a pre-communication phase via the subcarrier spreading technique [8].

The $\delta_{\nu_i}\left(R\left(\mathcal{N}_{U_k,i}\right)\right)$ of $\mathcal{N}_{U_k,i}$ parameters are determined from $\nu_i\left(\mathcal{N}_{U_k,i}\right)$ for all sub-channels via the iterative method of Theorem 2.

## S.1.2 Modulation Variance Adaption

The analysis focuses on a low-SNR CVQKD scenario. An initial low-SNR set of $\delta_{\nu_i}\left(R\left(\mathcal{N}_{U_k,i}\right)\right)$, $i = 0,\ldots,m-1$ of user $U_k$ from a low-SNR scenario is illustrated in Fig. S.1(a). The minimum of the set is $\xi_{U_k} = \min_{\forall i \in \mathcal{M}_{U_k}} \delta_{\nu_i}\left(R\left(\mathcal{N}_{U_k,i}\right)\right)$, from which the modulation variance correction for $\mathcal{N}_{U_k,i}$ is $\Delta_{\sigma^2_{\omega_i}} = \delta_{\nu_i}\left(R\left(\mathcal{N}_{U_k,i}\right)\right) - \xi_{U_k}$. The corresponding $\Delta_{\sigma^2_{\omega_i}}$ (see (58)), $i = 0,\ldots,m-1$ values determined from the $\delta_{\nu_i}\left(R\left(\mathcal{N}_{U_k,i}\right)\right)$ elements are depicted in Fig. S.1(b).

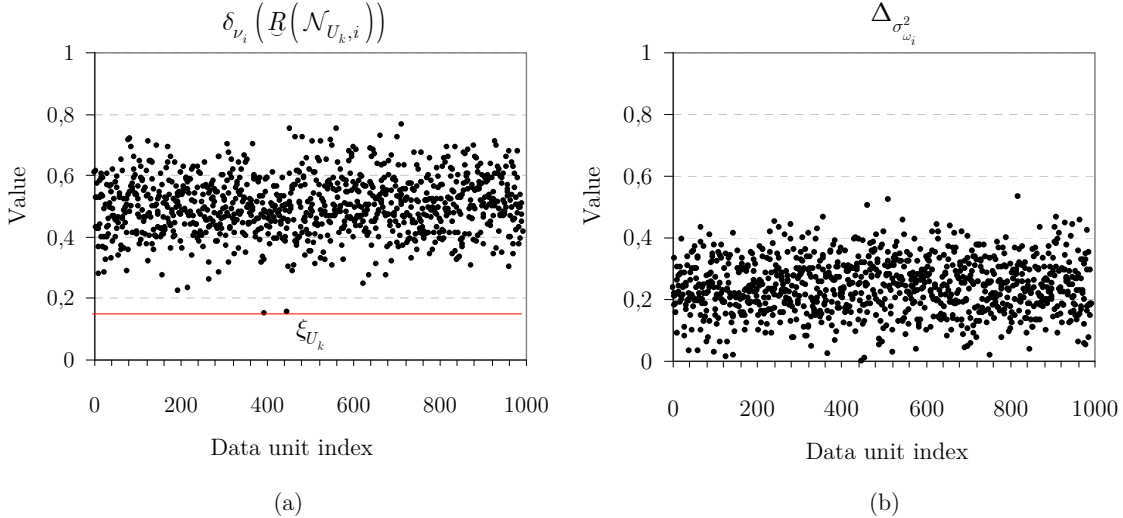

**Figure S.1.** An initial low-SNR set of $\delta_{\nu_i}\left(R\left(\mathcal{N}_{U_k,i}\right)\right)$ of user $U_k$, $i = 0,\ldots,m-1$, $m = 1000$. The minimum of set $\delta_{\nu_i}\left(R\left(\mathcal{N}_{U_k,i}\right)\right)$ is $\xi_{U_k}$ (red solid line) **(a)**. The resulting $\Delta_{\sigma^2_{\omega_i}}$ variance correction, $i = 0,\ldots,m-1$, $m = 1000$ determined from the $\delta_{\nu_i}\left(R\left(\mathcal{N}_{U_k,i}\right)\right)$ elements **(b)**.

In a low-SNR setting due to the value range of $\delta_{\nu_i}\left(R\left(\mathcal{N}_{U_k,i}\right)\right), i = 0,\ldots,m-1$, the required $\Delta_{\sigma^2_{\omega_i}}$ variance correction for the subcarriers is therefore is negligible.

The $x_{U_k,i}$, $i = 0,\ldots,m-1$ input quadratures of user $U_k$ with a constant variance $\sigma_\omega^2$ are illustrated in Fig. S.2(a). Applying the result of $\Delta_{\sigma_{\omega_i}^2}$, the $\tilde{x}_{U_k,i}$, $i = 0,\ldots,m-1$ input quadratures of user $U_k$ at the $\tilde{\sigma}_{\omega_i}^2 = \sigma_\omega^2 + \Delta_{\sigma_{\omega_i}^2}$ increased variance are depicted in Fig. S.2(b).

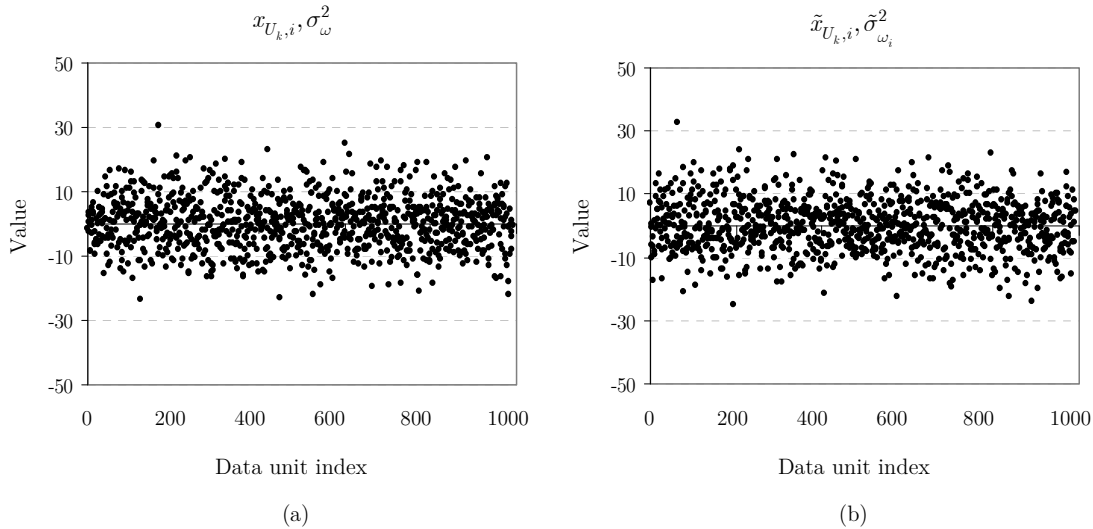

**Figure S.2.** The $x_{U_k,i}$, $i = 0,\ldots,m-1$, $m = 1000$ input quadratures of user $U_k$ at a constant variance $\sigma_\omega^2 = 64$ **(a)**. The $\tilde{x}_{U_k,i}$, $i = 0,\ldots,m-1$, $m = 1000$ input quadratures of user $U_k$ at increased variance $\tilde{\sigma}_{\omega_i}^2 = \sigma_\omega^2 + \Delta_{\sigma_{\omega_i}^2}$ **(b)**.

The variance adaption of $x_{U_k,i}$, $i = 0,\ldots,m-1$ requires only a moderate $\Delta_{\sigma_{\omega_i}^2}$ for all subcarriers to achieve an equalized, significantly lower error rate through the sub-channels.

## S.1.3 SNR Differences

The effects of the $\Delta_{\sigma_{\omega_i}^2}$ variance correction and the resulting $\varphi_{\delta_{\nu_i}}\left(R\left(\mathcal{N}_{U_k,i}\right)\right)$ can be expressed in terms of the resulting SNR change.

The $\Delta_{\mathrm{SNR}}\left(\tilde{\sigma}_{\omega_i}^2\right) = \mathrm{SNR}\left(\tilde{\sigma}_{\omega_i}^2\right) - \mathrm{SNR}\left(\sigma_\omega^2\right)$, $i = 0,\ldots,m-1$ SNR difference for the input quadratures $\left\{x_{U_k,i}, \tilde{x}_{U_k,i}\right\}$ at $\sigma_\omega^2$ and $\tilde{\sigma}_{\omega_i}^2 = \sigma_\omega^2 + \Delta_{\sigma_{\omega_i}^2}$ is depicted in Fig. S.3(a). The $\Delta_{\mathrm{SNR}}\left(\varphi_{\delta_{\nu_i}}\left(R\left(\mathcal{N}_{U_k,i}\right)\right)\right)$ SNR differences (see (61)) achieved at a given $\varphi_{\delta_{\nu_i}}\left(R\left(\mathcal{N}_{U_k,i}\right)\right)$, $i = 0,\ldots,m-1$ are illustrated in Fig. S.3(b).

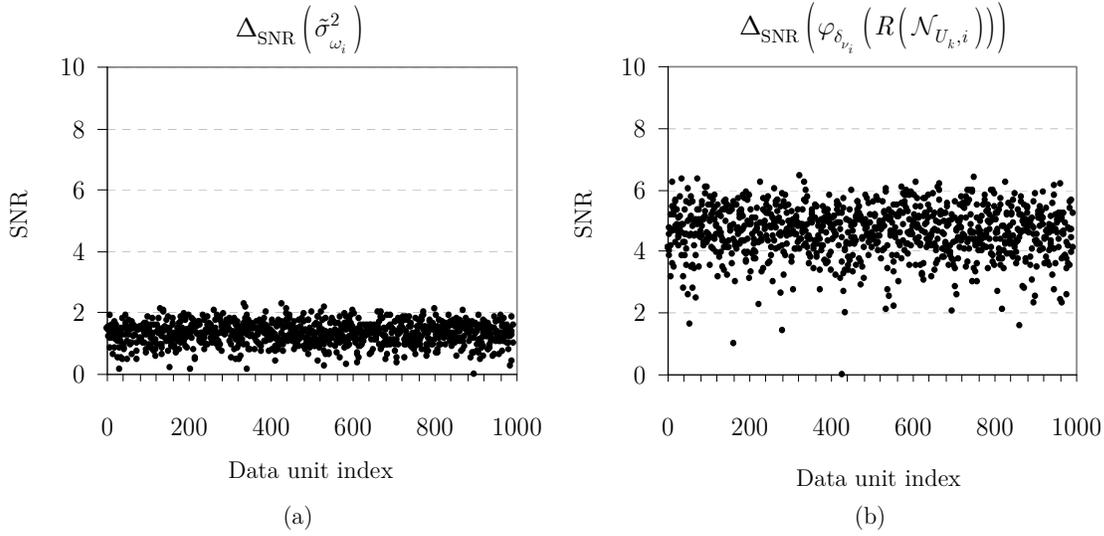

(a)   (b)

**Figure S.3.** The $\Delta_{\text{SNR}}\left(\tilde{\sigma}^2_{\omega_i}\right)$, $i = 0,\ldots,m-1$, $m = 1000$ parameter for $\left\{x_{U_k,i}, \tilde{x}_{U_k,i}\right\}$ at $\sigma^2_\omega$ and $\tilde{\sigma}^2_{\omega_i}$ **(a)**. The $\Delta_{\text{SNR}}\left(\varphi_{\delta_{\nu_i}}\left(R\left(\mathcal{N}_{U_k,i}\right)\right)\right)$ parameter, $i = 0,\ldots,m-1$, $m = 1000$ **(b)**.

The $\Delta_{\text{SNR}}\left(\tilde{\sigma}^2_{\omega_i}\right)$ input SNR difference results in an improved $\Delta_{\text{SNR}}\left(\varphi_{\delta_{\nu_i}}\left(R\left(\mathcal{N}_{U_k,i}\right)\right)\right)$ parameter via $\varphi_{\delta_{\nu_i}}\left(R\left(\mathcal{N}_{U_k,i}\right)\right)$ for all $i$.

## S.1.4  Error Rate Minimization

The BER values for the initial low-SNR set and for the set of $\varphi_{\delta_{\nu_i}}\left(R\left(\mathcal{N}_{U_k,i}\right)\right)$ are compared in Fig 4. The BER of $\mathcal{N}_{U_k,i}$ at the initial low-SNR set of $\delta_{\nu_i}\left(R\left(\mathcal{N}_{U_k,i}\right)\right)$, $i = 0,\ldots,m-1$ are illustrated in Fig. S.4(a). The BER of the $m$ sub-channels $\mathcal{N}_{U_k,i}$, $i = 0,\ldots,m-1$ at $\varphi_{\delta_{\nu_i}}\left(R\left(\mathcal{N}_{U_k,i}\right)\right)$, is depicted in Fig. S.4(b).

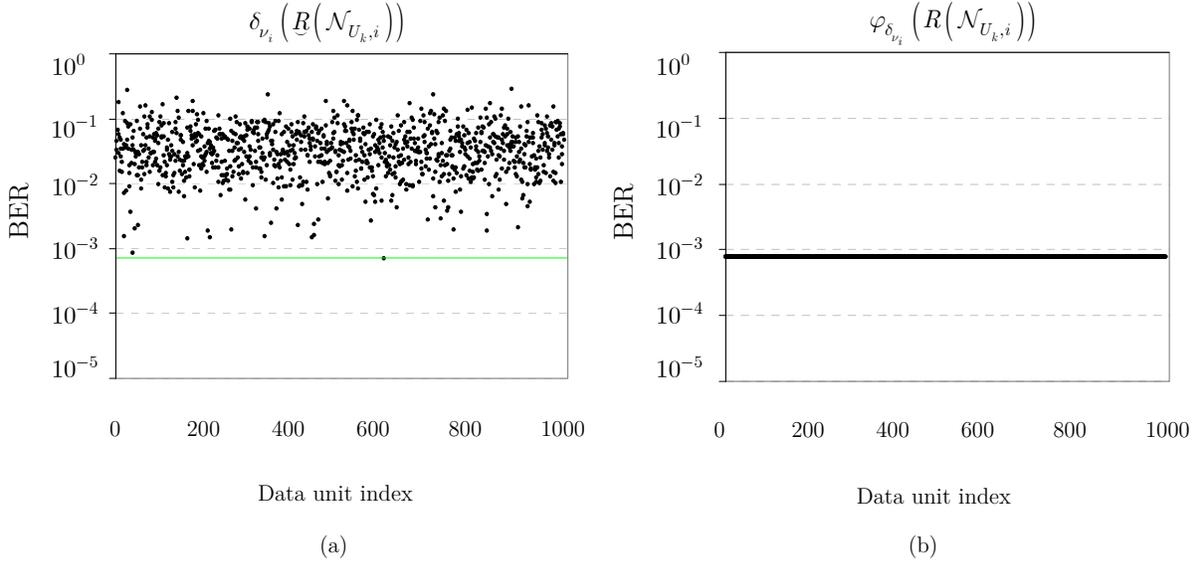

(a)  (b)

**Figure S.4.** The BER for the initial low-SNR set $\delta_{\nu_i}\left(\underline{R}\left(\mathcal{N}_{U_k,i}\right)\right)$, $i=0,\ldots,m-1$, $m=1000$ (the minimum of the set is depicted by the green solid line) **(a)**. The BER values for $\varphi_{\delta_{\nu_i}}\left(R\left(\mathcal{N}_{U_k,i}\right)\right)$, $i=0,\ldots,m-1$, $m=1000$ **(b)**.

As follows, as $\xi_{U_k}$ is determined and applied in the iteration procedure, the resulting BER is equally minimized for all $\mathcal{N}_{U_k,i}$ sub-channels of user $U_k$.

## S.2 Notations

The notations of the manuscript are summarized in Table S.1.

**Table S.1.** Summary of notations.

| Notation | Description |
|---|---|
| $i$ | Index for the $i$-th subcarrier Gaussian CV, $\left|\phi_i\right\rangle = x_i + \mathrm{i}p_i$. |
| $j$ | Index for the $j$-th Gaussian single-carrier CV, $\left|\varphi_j\right\rangle = x_j + \mathrm{i}p_j$. |
| $l$ | Number of Gaussian sub-channels $\mathcal{N}_i$ for the transmission of the Gaussian subcarriers. The overall number of the sub-channels is $n$. The remaining $n-l$ sub-channels do not transmit valuable information. |
| $x_i, p_i$ | Position and momentum quadratures of the $i$-th Gaussian subcarrier, $\left|\phi_i\right\rangle = x_i + \mathrm{i}p_i$. |
| $x'_i, p'_i$ | Noisy position and momentum quadratures of Bob's $i$-th noisy subcarrier Gaussian CV, $\left|\phi'_i\right\rangle = x'_i + \mathrm{i}p'_i$. |
| $x_j, p_j$ | Position and momentum quadratures of the $j$-th Gaussian single-carrier $\left|\varphi_j\right\rangle = x_j + \mathrm{i}p_j$. |
| $x'_j, p'_j$ | Noisy position and momentum quadratures of Bob's $j$-th recovered single-carrier Gaussian CV $\left|\varphi'_j\right\rangle = x'_j + \mathrm{i}p'_j$. |
| $x_{A,i}, p_{A,i}$ | Alice's quadratures in the transmission of the $i$-th subcarrier. |
| $\left|\phi_i\right\rangle, \left|\phi'_i\right\rangle$ | Transmitted and received Gaussian subcarriers. |
| $\mathbf{z} \in \mathcal{CN}\left(0, \mathbf{K}_\mathbf{z}\right)$ | A $d$-dimensional input CV vector to transmit valuable information. |
| $\mathbf{z}'^T$ | A $d$-dimensional noisy output vector, $\mathbf{z}'^T = \mathbf{A}^\dagger \mathbf{z} + \left(F^d\left(\Delta\right)\right)^T = \left(z'_0, \ldots, z'_{d-1}\right)$, where $z'_j = \left(\frac{1}{l}\sum_{i=0}^{l-1} F\left(T_{j,i}\left(\mathcal{N}_{j,i}\right)\right)\right)z_j + F\left(\Delta\right) \in \mathcal{CN}\left(0, 2\left(\sigma^2_{\omega_0} + \sigma^2_\mathcal{N}\right)\right)$. |
| $M$ | Measurement operator, homodyne or heterodyne measurement. |
| $\mathcal{S}$ | Set of private rates, identifies $r+2$ rate curves for a given sub-channel $\mathcal{N}_i$, as $\mathcal{S} = \left\{R_{\min}\left(\mathcal{N}_i\right), R\left(q\right), R_{\max}\left(\mathcal{N}_i\right)\right\}, q = 0\ldots r-1$, where $R_{\min}\left(\mathcal{N}_i\right) < R\left(0\right) \ldots < R\left(r-1\right) < R_{\max}\left(\mathcal{N}_i\right)$. |
| $R\left(q\right)$ | A target private rate, refers to a target transmission rate of private classical information over $\mathcal{N}_i$. Also identified by $R\left(\mathcal{N}_i\right)$. |
| $R\left(q-1\right)$ | A current private rate, refers to a current transmission rate of private classical information over $\mathcal{N}_i$. Also identified by $\underset{\sim}{R}\left(\mathcal{N}_i\right)$. |
| $P\left(\mathcal{N}_i\right)$ | Private classical capacity of a sub-channel $\mathcal{N}_i$. |
| $\chi_{AB}\left(\mathcal{N}_i\right)$ | Holevo information of Alice (transmitter) and Bob |
| | (receiver), with respect to sub-channel $\mathcal{N}_i$. |
| $\chi_{BE}(\mathcal{N}_i)$ | Holevo information of Bob and Eve (eavesdropper), with respect to sub-channel $\mathcal{N}_i$. |
| $S^*(\mathcal{N})$ | Target secret key rate $S^*(\mathcal{N}) = \sum_{i=0}^{l-1} R_i(\mathcal{N}_i)$, where $R_i(\mathcal{N}_i)$ is the private rate of sub-channel $\mathcal{N}_i$. |
| A | Adaption region, $A = [R_{\min}(\mathcal{N}_i), R_{\max}(\mathcal{N}_i)]$, contains $r$ private rate curves for the iteration. |
| $\mathrm{SNR}(R(\mathcal{N}_i))$ | SNR (signal to noise ratio) of $\mathcal{N}_i$ at private rate $R(\mathcal{N}_i)$, $$\mathrm{SNR}(R(\mathcal{N}_i)) = 10\log_{10}\frac{1}{\nu_i(R(\mathcal{N}_i))},$$ where $\nu_i(R(\mathcal{N}_i)) = \sigma^2_{R(\mathcal{N}_i)}\big/|F(T(R(\mathcal{N}_i)))|^2$, $\sigma^2_{R(\mathcal{N}_i)}$ is the noise variance of $\mathcal{N}_i$ at $R(\mathcal{N}_i)$, while $F(T(R(\mathcal{N}_i)))$ is the transmittance coefficient of $\mathcal{N}_i$ at $R(\mathcal{N}_i)$. |
| $\delta_{\nu_i}(R(\mathcal{N}_i))$ | Cumulative $\nu_i(R(\mathcal{N}_i))$ of $\mathcal{N}_i$ at an increased (target) private rate $R(\mathcal{N}_i)$, evaluated as $$\delta_{\nu_i}(R(\mathcal{N}_i)) = \delta_{\nu_i}(\underaccent{\smile}{R}(\mathcal{N}_i)) + \Delta_{\nu_i}(R(\mathcal{N}_i), \widehat{R}(\mathcal{N}_i)),$$ where $\underaccent{\smile}{R}(\mathcal{N}_i)$ is the current private rate, $R(\mathcal{N}_i)$ is the target private rate for $\mathcal{N}_i$, $\underaccent{\smile}{R}(\mathcal{N}_i) < R(\mathcal{N}_i) < \widehat{R}(\mathcal{N}_i)$. |
| $\Delta_{\nu_i}$ | Difference of $\nu_i$ at $R(\mathcal{N}_i)$ and $\widehat{R}(\mathcal{N}_i)$, $\widehat{R}(\mathcal{N}_i) > R(\mathcal{N}_i) > 0$, $$\Delta_{\nu_i} = \nu_i(R(\mathcal{N}_i)) - \nu_i(\widehat{R}(\mathcal{N}_i)).$$ |
| $\mathcal{B}$ | Bit error rate of $\mathcal{N}_i$. At target private rate $R(q)$ at a current rate $R(q-1)$, and $\delta_{\nu_i}(R(q-1))$, is expressed as $$\mathcal{B}\left(R(q)_{\delta_{\nu_i}(R(q-1))}\right).$$ |
| $\xi(R(q-1))$ | Identifies that $\mathcal{N}_i$, $i = 0,\ldots,l-1$ from the total $l$, for which $\delta_{\nu_i}(R(q-1))$ is minimal at a given $R(q)$, $$\xi(R(q-1)) = \min_{\forall i}\delta_{\nu_i}(R(q-1)),$$ where $\xi(0) = \min_{\forall i}\delta_{\nu_i}(0) = \min_{\forall i}\nu_i$. |
| $R(q)_{\delta_{\nu_i}(R(q-1))}$ | A target private rate $R(q)$ over $\mathcal{N}_i$ at a current $\delta_{\nu_i}(R(q-1))$. |
| $\mathcal{B}\left(R_{\min}(\mathcal{N}_i)_{\delta_{\nu_i}(R(q-1))}\right)$ | Minimal bit error rate, achievable at $R_{\min}(\mathcal{N}_i)$, at a given $\delta_{\nu_i}(R(q-1))$. |
| $\mathcal{F}(\cdot)$ | Function, evaluates SNR quantities from the input $\nu_i(\cdot)$ parameters, defined as $\mathcal{F}\big(\nu_i(\mathcal{N}_i)+\big(\Delta_{\nu_i}(R_{\min}(\mathcal{N}_i),R(0))+\ldots+\Delta_{\nu_i}(R(q-1),R(q))\big)\big)$ $=\left(\sqrt{\begin{array}{l}\mathrm{SNR}(\nu_i(\mathcal{N}_i))-[\mathrm{SNR}(\nu_i(R(0)))-\mathrm{SNR}(\nu_i(R_{\min}(\mathcal{N}_i)))]\\-\left(\sum_{k=0}^{q-1}[\mathrm{SNR}(\nu_i(R(k+1)))-\mathrm{SNR}(\nu_i(R(k)))]\right)\end{array}}\right)$, where $\mathrm{SNR}(\nu_i(R(x)))=10\log_{10}(1/(\nu_i(R(x))))$. |
| $erfc(\cdot)$ | Complementary error function, $erfc(x)=\frac{2}{\sqrt{\pi}}\int_x^\infty e^{-t^2}dt$. |
| $U_k$ | A given user, $k=0,\ldots,K-1$. |
| $\mathcal{M}_{U_k}$ | Logical channel of user $U_k, k=0,\ldots,K-1$, where $K$ is the number of total users, $$\mathcal{M}_{U_k}=\left[\mathcal{N}_{U_k,0},\ldots,\mathcal{N}_{U_k,m-1}\right]^T,$$ and $\mathcal{N}_{U_k,i}$ is the $i$-th sub-channel of $\mathcal{M}_{U_k}$, $m$ is the number of subcarriers dedicated to $U_k$. |
| $S^*(\mathcal{M}_{U_k})$ | Target secret key rate of $\mathcal{M}_{U_k}$, $$S^*(\mathcal{M}_{U_k})=\sum_{i=0}^{m-1}R(\mathcal{N}_{U_k,i}),$$ where $R(\mathcal{N}_{U_k,i})$ is the private rate of $\mathcal{N}_{U_k,i}$. |
| $\underline{R}(\mathcal{N}_{U_k,i})$ | A current private rate of $\mathcal{N}_{U_k,i}$, $\underline{R}(\mathcal{N}_{U_k,i})=R_i(q-1)$. |
| $R(\mathcal{N}_{U_k,i})$ | A target private rate of $\mathcal{N}_{U_k,i}$, $R(\mathcal{N}_{U_k,i})=R_i(q)$ where $R_i(q)$ refers to the $R(q)$ rate of $\mathcal{N}_{U_k,i}$, with relation $R_i(q-1)<R_i(q)$, $R_i(-1)=R_{\min}(\mathcal{N}_{U_k,i})$, and $R_i(r)=R_{\max}(\mathcal{N}_{U_k,i})$. |
| $\delta_{\nu_i}(R_i(q))$ | Parameter $\delta_{\nu_i}$ of $\mathcal{N}_{U_k,i}$ at private rate $R(\mathcal{N}_{U_k,i})$. |

| Symbol | Description |
|---|---|
| $\xi_{U_k}$ | Identifies the minimal $\delta_{\nu_i}\left(R\left(\mathcal{N}_{U_k,i}\right)\right)$ parameter for the $\mathcal{N}_{U_k,i}$, $i=0,\ldots,m-1$ sub-channels of the set $\mathcal{M}_{U_k}$, as $$\xi_{U_k} = \min_{\forall i \in \mathcal{M}_{U_k}} \delta_{\nu_i}\left(R\left(\mathcal{N}_{U_k,i}\right)\right).$$ |
| $\Delta_{\sigma^2_{\omega_i}}$ | Modulation variance adaption at rate $R\left(\mathcal{N}_{U_k,i}\right)$, $$\Delta_{\sigma^2_{\omega_i}} = \delta_{\nu_i}\left(R\left(\mathcal{N}_{U_k,i}\right)\right) - \xi_{U_k}.$$ |
| $\tilde{\sigma}^2_\omega$ | A corrected modulation variance, $\tilde{\sigma}^2_\omega = \sigma^2_\omega + \Delta_{\sigma^2_{\omega_i}}$, where $\sigma^2_\omega$ is the initial input subcarrier modulation variance, and $\Delta_{\sigma^2_{\omega_i}} = \delta_{\nu_i}\left(R\left(\mathcal{N}_{U_k,i}\right)\right) - \xi_{U_k}$. |
| $\varphi_{\delta_{\nu_i}}\left(R\left(\mathcal{N}_{U_k,i}\right)\right)$ | The resulting $\delta_{\nu_i}$ parameter for $\mathcal{N}_{U_k,i}$ at a target rate $R\left(\mathcal{N}_{U_k,i}\right)$, $\varphi_{\delta_{\nu_i}}\left(R\left(\mathcal{N}_{U_k,i}\right)\right) = \delta_{\nu_i}\left(R\left(\mathcal{N}_{U_k,i}\right)\right) - \xi_{U_k}$. |
| $\Delta_{\text{SNR}}\left(\varphi_{\delta_{\nu_i}}\left(R\left(\mathcal{N}_{U_k,i}\right)\right)\right)$ | SNR increment of $\mathcal{N}_{U_k,i}$, achieved by $\Delta_{\sigma^2_{\omega_i}}$ at a given $R\left(\mathcal{N}_{U_k,i}\right) = R(q)$. |
| $\mathcal{B}\left(R\left(\mathcal{N}_{U_k,i}\right)_{\xi_{U_k}}\right)$ | An equally minimized bit error rate for all $\mathcal{N}_{U_k,i}$ at an arbitrary $R\left(\mathcal{N}_{U_k,i}\right)_{\delta_{\nu_i}\left(R\left(\mathcal{N}_{U_k,i}\right)\right)}$. |
| $F$ | Fourier transform (FFT). |
| $\vec{\varphi}_{U_k}$ | A $d$-dimensional input vector of user $U_k$, $k=0,\ldots,K-1$, $\vec{\varphi}_{U_k} = \left(\varphi_{U_k,0},\ldots,\varphi_{U_k,d-1}\right)^T$, where $\varphi_{U_k,j}$ refers to the $j$-th single-carrier CV, $\varphi_{U_k,j} = x_{U_k,j} + \mathrm{i}p_{U_k,j}$, and $\left\{x_{U_k,j}, p_{U_k,j}\right\}$ are Gaussian random quadratures. |
| $\vec{x}_{U_k}$ | A $d$-dimensional vector of $U_k$, $\vec{x}_{U_k} = \left(x_{U_k,0},\ldots,x_{U_k,d-1}\right)^T$, where $x_{U_k,j}$ is the quadrature component of $\varphi_{U_k,j}$. |
| $x'_{U_k,i}$ | Noisy subcarrier, discrete variable, $x'_{U_k,i} = M\left(\phi'_{U_k,i}\right)$, where $M$ is a measurement operator, and $\phi'_{U_k,i}$ is the $i$-th noisy Gaussian subcarrier CV of $U_k$. |
| $\mathbf{x}_{U_k,i}$ | Input subcarrier vector, $\mathbf{x}_{U_k,i} = \left(x_{U_k,0},\ldots,x_{U_k,m-1}\right)^T$. |
| $\mathbf{x}'_{U_k,i}$ | Output subcarrier vector, $\mathbf{x}'_{U_k,i} = \left(x'_{U_k,0},...,x'_{U_k,m-1}\right)^T$. |
| $z \in \mathcal{CN}\left(0,\sigma_z^2\right)$ | The variable of a single-carrier Gaussian CV state, $\left|\varphi_i\right\rangle \in \mathcal{S}$. Zero-mean, circular symmetric complex Gaussian random variable, $\sigma_z^2 = \mathbb{E}\left[\left|z\right|^2\right] = 2\sigma_{\omega_0}^2$, with i.i.d. zero mean, Gaussian random quadrature components $x,p \in \mathbb{N}\left(0,\sigma_{\omega_0}^2\right)$, where $\sigma_{\omega_0}^2$ is the variance. |
| $\Delta \in \mathcal{CN}\left(0,\sigma_\Delta^2\right)$ | The noise variable of the Gaussian channel $\mathcal{N}$, with i.i.d. zero-mean, Gaussian random noise components on the position and momentum quadratures $\Delta_x, \Delta_p \in \mathbb{N}\left(0,\sigma_\mathcal{N}^2\right)$, $\sigma_\Delta^2 = \mathbb{E}\left[\left|\Delta\right|^2\right] = 2\sigma_\mathcal{N}^2$. |
| $d \in \mathcal{CN}\left(0,\sigma_d^2\right)$ | The variable of a Gaussian subcarrier CV state, $\left|\phi_i\right\rangle \in \mathcal{S}$. Zero-mean, circular symmetric Gaussian random variable, $\sigma_d^2 = \mathbb{E}\left[\left|d\right|^2\right] = 2\sigma_\omega^2$, with i.i.d. zero mean, Gaussian random quadrature components $x_d, p_d \in \mathbb{N}\left(0,\sigma_\omega^2\right)$, where $\sigma_\omega^2$ is the (constant) modulation variance of the Gaussian subcarrier CV state. |
| $F^{-1}\left(\cdot\right) = \text{CVQFT}^\dagger\left(\cdot\right)$ | The inverse CVQFT transformation, applied by the encoder, continuous-variable unitary operation. |
| $F\left(\cdot\right) = \text{CVQFT}\left(\cdot\right)$ | The CVQFT transformation, applied by the decoder, continuous-variable unitary operation. |
| $F^{-1}\left(\cdot\right) = \text{IFFT}\left(\cdot\right)$ | Inverse FFT transform, applied by the encoder. |
| $\sigma_{\omega_0}^2$ | Single-carrier modulation variance. |
| $\sigma_\omega^2 = \frac{1}{l}\sum_l \sigma_{\omega_i}^2$ | Multicarrier modulation variance. Average modulation variance of the $l$ Gaussian sub-channels $\mathcal{N}_i$. |
| $\left|\phi_i\right\rangle = \left|\text{IFFT}\left(z_{k,i}\right)\right\rangle$ $= \left|F^{-1}\left(z_{k,i}\right)\right\rangle = \left|d_i\right\rangle.$ | The $i$-th Gaussian subcarrier CV of user $U_k$, where IFFT stands for the Inverse Fast Fourier Transform, $\left|\phi_i\right\rangle \in \mathcal{S}$, $d_i \in \mathcal{CN}\left(0,\sigma_{d_i}^2\right)$, $\sigma_{d_i}^2 = \mathbb{E}\left[\left|d_i\right|^2\right]$, $d_i = x_{d_i} + \mathrm{i}p_{d_i}$, $x_{d_i} \in \mathbb{N}\left(0,\sigma_{\omega_F}^2\right)$, $p_{d_i} \in \mathbb{N}\left(0,\sigma_{\omega_F}^2\right)$ are i.i.d. zero-mean |

| | |
|---|---|
| | Gaussian random quadrature components, and $\sigma^2_{\omega_F}$ is the variance of the Fourier transformed Gaussian state. |
| $\left|\varphi_{k,i}\right\rangle = \mathrm{CVQFT}\left(\left|\phi_i\right\rangle\right)$ | The decoded single-carrier CV of user $U_k$ from the subcarrier CV, expressed as $F\left(\left|d_i\right\rangle\right) = \left|F\left(F^{-1}\left(z_{k,i}\right)\right)\right\rangle = \left|z_{k,i}\right\rangle$. |
| $\mathcal{N}$ | Gaussian quantum channel. |
| $\mathcal{N}_i, i = 0,...,n-1$ | Gaussian sub-channels. |
| $T(\mathcal{N})$ | Channel transmittance, normalized complex random variable, $T(\mathcal{N}) = \operatorname{Re} T(\mathcal{N}) + \mathrm{i} \operatorname{Im} T(\mathcal{N}) \in \mathcal{C}$. The real part identifies the position quadrature transmission, the imaginary part identifies the transmittance of the position quadrature. |
| $T_i(\mathcal{N}_i)$ | Transmittance coefficient of Gaussian sub-channel $\mathcal{N}_i$, $T_i(\mathcal{N}_i) = \operatorname{Re}\left(T_i(\mathcal{N}_i)\right) + \mathrm{i} \operatorname{Im}\left(T_i(\mathcal{N}_i)\right) \in \mathcal{C}$, quantifies the position and momentum quadrature transmission, with (normalized) real and imaginary parts $0 \leq \operatorname{Re} T_i(\mathcal{N}_i) \leq 1/\sqrt{2}$, $0 \leq \operatorname{Im} T_i(\mathcal{N}_i) \leq 1/\sqrt{2}$, where $\operatorname{Re} T_i(\mathcal{N}_i) = \operatorname{Im} T_i(\mathcal{N}_i)$. |
| $T_{Eve}$ | Eve's transmittance, $T_{Eve} = 1 - T(\mathcal{N})$. |
| $T_{Eve,i}$ | Eve's transmittance for the $i$-th subcarrier CV. |
| $\mathbf{z} = \mathbf{x} + \mathrm{i}\mathbf{p} = \left(z_0,...,z_{d-1}\right)^T$ | A $d$-dimensional, zero-mean, circular symmetric complex random Gaussian vector that models $d$ Gaussian CV input states, $\mathcal{CN}(0,\mathbf{K_z})$, $\mathbf{K_z} = \mathbb{E}\left[\mathbf{z}\mathbf{z}^\dagger\right]$, where $z_j = x_j + \mathrm{i}p_j$, $\mathbf{x} = \left(x_0,...,x_{d-1}\right)^T$, $\mathbf{p} = \left(p_0,...,p_{d-1}\right)^T$, $x_j \in \mathbb{N}\left(0,\sigma^2_{\omega_0}\right)$, $p_j \in \mathbb{N}\left(0,\sigma^2_{\omega_0}\right)$ i.i.d. zero-mean Gaussian random variables. |
| $\mathbf{d} = F^{-1}(\mathbf{z})$ | An $l$-dimensional, zero-mean, circular symmetric complex random Gaussian vector, $\mathcal{CN}(0,\mathbf{K_d})$, $\mathbf{K_d} = \mathbb{E}\left[\mathbf{d}\mathbf{d}^\dagger\right]$, $\mathbf{d} = \left(d_0,...,d_{l-1}\right)^T, d_i = x_i + \mathrm{i}p_i$, $x_i, p_i \in \mathbb{N}\left(0,\sigma^2_{\omega_F}\right)$ are i.i.d. zero-mean Gaussian random variables. The $i$-th component is $d_i \in \mathcal{CN}\left(0,\sigma^2_{d_i}\right)$, $\sigma^2_{d_i} = \mathbb{E}\left[\left|d_i\right|^2\right]$. |

| | |
|---|---|
| $\mathbf{y}_k \in \mathcal{CN}\left(0, \mathbb{E}\left[\mathbf{y}_k \mathbf{y}_k^\dagger\right]\right)$ | A $d$-dimensional zero-mean, circular symmetric complex Gaussian random vector. |
| $y_{k,m}$ | The $m$-th element of the $k$-th user's vector $\mathbf{y}_k$, expressed as $y_{k,m} = \sum_l F\left(T_i\left(\mathcal{N}_i\right)\right) F\left(d_i\right) + F\left(\Delta_i\right)$. |
| $F\left(\mathbf{T}\left(\mathcal{N}\right)\right)$ | Fourier transform of $\mathbf{T}\left(\mathcal{N}\right) = \left[T_0\left(\mathcal{N}_0\right)..., T_{l-1}\left(\mathcal{N}_{l-1}\right)\right]^T \in \mathcal{C}^l$, the complex transmittance vector. |
| $F\left(\Delta\right)$ | Complex vector, expressed as $F\left(\Delta\right) = e^{\frac{-F(\Delta)^T \mathbf{K}_{F(\Delta)} F(\Delta)}{2}}$, with covariance matrix $\mathbf{K}_{F(\Delta)} = \mathbb{E}\left[F\left(\Delta\right) F\left(\Delta\right)^\dagger\right]$. |
| $\mathbf{y}[j]$ | AMQD block, $\mathbf{y}[j] = F\left(\mathbf{T}\left(\mathcal{N}\right)\right) F\left(\mathbf{d}\right)[j] + F\left(\Delta\right)[j]$. |
| $\tau = \|F\left(\mathbf{d}\right)[j]\|^2$ | An exponentially distributed variable, with density $f\left(\tau\right) = \left(1/2\sigma_\omega^{2n}\right) e^{-\tau/2\sigma_\omega^2}$, $\mathbb{E}[\tau] \leq n 2\sigma_\omega^2$. |
| $T_{Eve,i}$ | Eve's transmittance on the Gaussian sub-channel $\mathcal{N}_i$, $T_{Eve,i} = \operatorname{Re} T_{Eve,i} + i \operatorname{Im} T_{Eve,i} \in \mathcal{C}$, $0 \leq \operatorname{Re} T_{Eve,i} \leq 1/\sqrt{2}$, $0 \leq \operatorname{Im} T_{Eve,i} \leq 1/\sqrt{2}$, $0 \leq |T_{Eve,i}|^2 < 1$. |
| $d_i$ | A $d_i$ subcarrier in an AMQD block. |
| $\nu_{\min}$ | The $\min\{\nu_0,...,\nu_{l-1}\}$ minimum of the $\nu_i$ sub-channel coefficients, where $\nu_i = \sigma_\mathcal{N}^2 / \left|F\left(T_i\left(\mathcal{N}_i\right)\right)\right|^2$ and $\nu_i < \nu_{Eve}$. |
| $\sigma_\omega^2$ | Constant modulation variance, $\sigma_\omega^2 = \nu_{Eve} - \nu_{\min} \mathcal{G}(\delta)_{p(x)}$, $\nu_{Eve} = \frac{1}{\lambda}$, $\lambda = \left|F\left(T_\mathcal{N}^*\right)\right|^2 = \frac{1}{n} \sum_{i=0}^{n-1} \left|\sum_{k=0}^{n-1} T_k^* e^{\frac{-i2\pi ik}{n}}\right|^2$ and $T_\mathcal{N}^*$ is the expected transmittance of the Gaussian sub-channels under an optimal Gaussian collective attack. |

## S.3 Abbreviations

| | |
|---|---|
| **AMQD** | **Adaptive Multicarrier Quadrature Division** |
| **AWGN** | **Additive White Gaussian Noise** |
| **CV** | **Continuous-Variable** |
| **CVQFT** | **Continuous-Variable Quantum Fourier Transform** |
| **CVQKD** | **Continuous-Variable Quantum Key Distribution** |
| **DV** | **Discrete Variable** |
| **FFT** | **Fast Fourier Transform** |
| **ICVQFT** | **Inverse CVQFT** |
| **IFFT** | **Inverse Fast Fourier Transform** |
| **MQA** | **Multiuser Quadrature Allocation** |
| **QFT** | **Quantum Fourier Transform** |
| **QKD** | **Quantum Key Distribution** |
| **SNR** | **Signal to Noise Ratio** |